\documentclass[11pt]{article}

\usepackage{sectsty}
\usepackage{graphicx}
\usepackage{natbib}
\usepackage{booktabs}
\usepackage{amsmath}
\usepackage{amsfonts}
\usepackage{amssymb}
\usepackage{rotating}
\usepackage{hyperref}

\def\sidewaystablefn{\renewcommand\footnotetext[2][]{{\removelastskip\vskip3pt%
\let\tablebodyfont\tablefootnotefont%
\hskip0pt\if!##1!\else{\smash{$^{##1}$}}\fi##2\par}}%
}%

\begin{document}

\title{Consumer-side Fairness in Recommender Systems: A Systematic Survey of Methods and Evaluation}
\author{Bjørnar Vassøy and Helge Langseth\\ Norwegian University of Science and Technology }
\date{}
\maketitle





\abstract{In the current landscape of ever-increasing levels of digitalization, we are facing major challenges pertaining to scalability. Recommender systems have become irreplaceable both for helping users navigate the increasing amounts of data and, conversely, aiding providers in marketing products to interested users. The growing awareness of discrimination in machine learning methods has recently motivated both academia and industry to research how fairness can be ensured in recommender systems. For recommender systems, such issues are well exemplified by occupation recommendation, where biases in historical data may lead to recommender systems relating one gender to lower wages or to the propagation of stereotypes. In particular, consumer-side fairness, which focuses on mitigating discrimination experienced by users of recommender systems, has seen a vast number of diverse approaches for addressing different types of discrimination. The nature of said discrimination depends on the setting and the applied fairness interpretation, of which there are many variations. This survey serves as a systematic overview and discussion of the current research on consumer-side fairness in recommender systems. To that end, a novel taxonomy based on high-level fairness interpretation is proposed and used to categorize the research and their proposed fairness evaluation metrics. Finally, we highlight some suggestions for the future direction of the field.}


\newpage

\section{Introduction}
Recommender systems have become integral parts of modern digital society. An exponential increase in data poses significant challenges to users and consumers, who cannot feasibly sift through everything to find what they are looking for. Recommender systems help mitigate these challenges by capturing their users' preferences and presenting them with prioritized options. Thus, recommender systems have seen widespread application in e-commerce, multimedia platforms, and social networks. Their tactical relevance in the industry has led to a high degree of cooperation between the industry and academia in further developing the field.

In recent years, the notion of fairness in machine learning has steadily gained attention. High-profile cases have succeeded in bringing the topic to the general public's attention, like the analysis performed by ProPublica suggesting the presence of racial bias in the COMPAS system used for predicting the likelihood of recidivism of inmates \citep{propublica_2016}. Subsequently, fairness challenges have also been identified for recommender systems, and the works of \cite{multi_burke_2017} formalized the presence of multi-stakeholder fairness dynamics mirroring the multi-stakeholder nature of recommender systems. Provider stakeholders may take issue if their products are disproportionally less exposed than similar popular products. Career seekers may feel discriminated against if they are predominantly recommended careers that are stereotypically and historically associated with their gender. An increased focus on fairness in recommender systems is not only ethically beneficial for society as a whole but also helps the actors applying them in satisfying an increasingly fairness-aware user base and retaining good relations and cooperation with providers. 

While provider-side fairness research has a dominant subgroup in research pursuing popularity bias, which is the notion of disproportional amounts of attention given to popular items, consumer-side fairness research has a greater focus on group-based fairness relating to demographic information of the users, i.e., making sure that users are not discriminated against based on aspects such as race, gender, or age. Despite the focus on a specific high-level fairness setting, consumer-side fairness in recommender systems displays a high degree of variation in approaches. The approaches for introducing fairness awareness take place in all parts of the recommender system pipeline, span most established and upcoming model architectures, and are designed for respecting various fairness interpretations. Some models opt for adjusting recommendations post hoc, others modify the input data directly, while others still explicitly model the fairness-awareness. Fairness has been incorporated through penalizing discrimination during optimization, altering user representation to be more neutral, probabilistically modelling the influence of sensitive attributes, or re-ranking unaltered recommendation, all while adhering to different definitions of what discrimination and fairness entails. There is also variation in the application setting of these approaches; most adhere to the regular asymmetric setting where users and items make up fundamentally different concepts, while others consider reciprocal settings where users are recommended to other users like matchmaking. Yet another dynamic is considered in two-sided settings that seek to achieve both consumer- and provider-side fairness concurrently. Despite the great variety,  the breadth of consumer-side fairness approaches has yet to be covered in detail by any existing surveys. We further argue for this claim in Section \ref{sec:related_work}, where we discuss relevant surveys.

In this survey, we have systematically surveyed the existing literature that proposes and evaluates approaches considering consumer-side fairness. Critical aspects of the qualified literature are discussed, compared, and categorized, leading to the proposal of a taxonomy that highlights fairness interpretation and how it has been incorporated into the approaches. Further, we provide a comprehensive overview of metrics used to evaluate the fairness of the approaches and some thoughts on the field's future directions. Our key contributions are:
\begin{enumerate}
    \item Propose a taxonomy for categorizing consumer-side fairness approaches in recommender systems based on how the fairness is incorporated and high-level conceptual fairness definitions.
    \item Provide a comprehensive overview, categorization, and comparison of available consumer-side fairness approaches in recommender systems and their proposed fairness evaluation.
\end{enumerate}

The remaining sections of this survey include a background section on fairness definitions, terminology, related concepts, and related works; methodology covering the literature selection process and the proposed taxonomy; a detailed discussion and comparison of the identified literature; analysis of applied fairness metrics and datasets; and a final discussion of our thoughts on the future directions of the topic. 

\section{Background}
As a primer to this survey's core content and discussion, we introduce key established fairness concepts and terms that appear frequently or are subject to ambiguity. The background also covers a discussion of recommender systems concepts related to consumer-side fairness and a look into existing surveys on fairness in recommender systems and how this survey differs. 

\subsection{Terminology}\label{sec:terminology}
The following definitions have been added to mitigate confusion stemming from mixing similar terms or different interpretations of specific terms. A low degree of consensus, especially within fairness-focused research, has led to multiple different terms being used for the same concept and other words like \textit{preference} are contextually ambiguous.

\noindent\textbf{Rating:} In rating-based recommender systems, we are interested in the rating given by a specific user to a specific item and is contrasted with ranking-based recommender systems. Ratings can be discrete and continuous and typically have a set range, e.g., between 1 and 5.

\noindent\textbf{Ranking:} Ordering of items or entities according to users' (perceived) preference.

\noindent\textbf{IR Ranking:} The field of Information Retrieval comprise an array of different approaches for retrieving information from data storage. We will consider intent the key factor separating IR Ranking and recommender systems: recommender systems seek to suggest novel, but relevant, information to their users while IR Ranking seeks to retrieve the most relevant information. Furthermore, IR Ranking approaches often involve a query and are rarely personalized.

\noindent\textbf{Top-$k$:} Top $k$ ranked item, where $k$ is an integer indicating the number of items that are of interest. $k$ is usually quite small, often in the range of 5-20, as user attention is a limiting factor. 

\noindent\textbf{Preference (score):} Continuous measure of user preference used to produce rankings. \\Score/value/measure may be omitted in the text if the context allows it.

\noindent\textbf{Ranking-based recommender systems:} Recommender systems that learn to rank in order to present the user with the top list of suggested items. Usually applies classification-based optimization.

\noindent\textbf{Rating-based recommender systems:} Recommender systems that attempt to match ratings given to items by users, and predict new ratings given by users to unrated items. Usually applies regression-based optimization.

\noindent\textbf{Sensitive attribute:} Unifying term used to describe demographic attributes that are used to segment users into different groups for which fairness considerations are applied. Similar concepts, both symmetric and asymmetric, have been referred to as \textit{demographic}, \textit{protected}, \textit{minority}, \textit{marginalized} and \textit{private} in the selected studies. \textit{Sensitive attribute} is found to be sufficient for explaining most approaches, but more thorough explanations are provided in cases where asymmetry or special dynamics of a sensitive attribute take a more nuanced role.

\noindent\textbf{Sensitive group:} A group of users that share a specific instance of a sensitive attribute, e.g., all male users in a setting where gender is considered a sensitive attribute.

\subsection{Recommender System Definition}\label{sec:rec_sys_def}
\begin{table}[htb]
\begin{center}
\caption{Notation}\label{tab:basic_notation}%
\begin{tabular}{ p{0.08\textwidth} p{0.85\textwidth}}
\toprule
\textbf{Symbols} & \textbf{Description}\\
\midrule
$\mathcal{U}$&Set of all users\\
$\mathcal{V}$&Set of all items/recommendable entities\\
$\mathcal{S}$&Set of all possible sensitive attribute configurations\\
$r$&Rating in rating-based recommender systems. Double as preference in mixed usage\\
$\text{pref}$&Intermediate measure of preference used in ranking-based recommender systems\\
$y$&Binary indicator for the presence of recommendation in ranking-based recommender systems\\
$\hat{ }$&Modifier, indicate predicted output as opposed to ground truth\\
$_{u,v,s}$&Indicate relation with specific users, items or sensitive values respectively\\
$\text{Rec}$&Set of (top-$k$) recommendations\\
$\text{Util}(\cdot)$&Open-ended/arbitrary utility function\\
$\text{P}(\cdot)$&Probability\\
$\hat{\mathbb{E}}[\cdot]$&  Arithmetic Mean\\
\bottomrule
\end{tabular}
\end{center}
\end{table}
Recommender systems comprise many varied approaches designed for varied settings and present no clear singular definition. We will focus on personalized recommender systems, i.e., those that seek to accommodate different individuals with customized recommendations based on their individual preferences. As alluded to in Section \ref{sec:terminology}, this survey will distinguish between rating-based and ranking-based recommender systems. When applying the notation presented in Table \ref{tab:basic_notation}, both flavours attempt to capture how a set of entities $\mathcal{U}$ will value another set of entities $\mathcal{V}$ on an individual level. $\mathcal{U}$ are typically exemplified as \textit{users} and $\mathcal{V}$ as \textit{items}, and the overall goal is to \textit{recommend} novel items to the users. For rating-based recommender systems, the level objective is to predict individual ratings given by a user $u$ to an item $v$, $r_{uv}$, i.e., $\hat{r}_{uv} = r_{uv}$. Ranking-based recommender systems instead take the approach of capturing the general preferences of the users and using this to present the same users with selections of items predicted to be of the users' liking. The resulting objective is analogous to the rating-based objective, $\hat{y}_{uv} = y_{uv}$, but does present slightly different challenges owing to the non-continuous nature of ranking. Both Rating-based and Ranking-based recommender systems may adapt rating data, but Ranking-based recommender systems can more easily adapt data of a more implicit nature, e.g., interaction events.

Recommender systems are implemented using a plethora of different models and methods like Neighbourhood-based Collaborative Filtering \citep{cf_2022}, Matrix Factorization \citep{mf_2022}, various types of Deep Neural Networks \citep{deep_rec_2022}, Autoencoders \citep{vae_rec_2017}, Reinforcement Learning \citep{reinforce_2021}, Graph-based models \citep{graph_learning_2021}, and various Probabilistic models. Detailed background theories of various models have been left out to avoid significant inflation of the survey's length. However, as this is a comprehensive survey focusing on tangible model proposals, some technical details will be discussed. Readers are encouraged to consult auxiliary sources, like the provided references, when needed. 

\subsection{Formal Fairness Definitions}\label{sec:formal}
Several formal fairness definitions have been proposed for classification settings. While some of these can be trivially adapted to the recommendation setting, others are more challenging. One such challenge relates to adaptations of definitions based on confusion matrix statistics, as the interpretations and implications of these statistics may differ between classification and recommender systems. Confusion matrixes are not computable nor relevant for rating-based recommender systems. Conversely, the confusion matrixes of ranking-based recommender systems are heavily influenced by the fixed number of recommendations and the number of correct recommendations, which usually vary by user. Furthermore, the implications of some definitions may be enhanced in scenarios where a positive label is deemed a positive outcome for a stakeholder, even if it was a False Positive. An example of this could be an applicant applying for a loan; the applicant will be happy if the application is accepted regardless of whether it was the correct verdict according to bank policies. In consumer-side recommendation settings, this is rarely true, in which case, a False Positive in a top-$k$ recommendation setting will simply be the presence of an item that the user does not care for among the top recommendations.

A selection of fairness definitions is covered here, along with accompanying descriptions of recommender system-specific adaptions. The reader is encouraged to consult \cite{formalizing_fair_2017}, \cite{fair_ml_survey_2020}, and \cite{survey_li_2022} for a more in-depth discussion of formal fairness definitions in both machine learning and recommender systems.

\subsubsection{Fairness Through Unawareness}
One of the more naive fairness definitions is achieved by simply omitting explicit sensitive attributes in the modelling. This definition is widely disregarded as it fails to consider implicit biases present in other attributes and is therefore not sufficient for mitigating discrimination \citep{formalizing_fair_2017}.

\subsubsection{Statistical Parity}
Statistical parity in classification requires that each group has an equal probability of being assigned a positive label.
$$\text{P}(\hat{y}=1\arrowvert s=s_{1}) = \text{P}(\hat{y}=1\arrowvert s=s_{2}) = ...$$
Where $\text{P}(\cdot)$ represents probability, $\hat{y}$ is the predicted label, and $s$ is a sensitive attribute.

Statistical Parity is adaptable in recommender systems by replacing the notion of classification labels with ratings or discrete recommended items, but its evaluation may be trickier.

\subsubsection{Equal Opportunity}
Equal opportunity in classification requires that the true positive rate of different sensitive groups is equal. 
\begin{equation}
    \text{P}(\hat{y} = 1\arrowvert y = 1, s = s_1) = \text{P}(\hat{y} = 1\arrowvert y = 1, s = s_2) = ...\nonumber
\end{equation}

\subsubsection{Equalized Odds}
The Equalized Odds definition is stricter than Equal Opportunity in also requiring that the false positive rates of the different sensitive groups are equal. 
\begin{align}
    &\text{P}(\hat{y} = 1\arrowvert y = 1, s = s_1) = \text{P}(\hat{y} = 1\arrowvert y = 1, s = s_2) = ...\nonumber\\
    \&\ &\text{P}(\hat{y} = 1\arrowvert y = 0, s = s_1) = \text{P}(\hat{y} = 1\arrowvert y = 0, s = s_2) = ...\nonumber
\end{align}
As previously mentioned, false positives may benefit some stakeholders in certain scenarios. However, false positives may also be the most detrimental type of error in other scenarios. Thus, the decision to pursue Equalized Odds instead of Equal Opportunity may be motivated by a wish to balance either a boon or a bane. 

\subsection{Related Recommender System Concepts}\label{sec:rel_rec_conc}
Many flavours of recommender systems have arisen to cover different needs as they  appeared or were made known. It is not uncommon that the concepts considered in these different flavours partly overlap, share underlying issues, or share similar mitigation strategies. A number of the most relevant recommender system flavours, when compared with consumer-side fairness, are listed in this section, focusing on similarities and dissimilarities. The intention of this section is two-fold: The first is to highlight related topics that may be of interest to readers and that may help put consumer-side fairness into a broader context. The second motive is to highlight dissimilarities that disqualify certain research from being covered by the scope of this survey, despite occasionally adopting fairness terminology. 

\subsubsection{Provider-side fairness}
As the name entails, provider-side fairness takes a polar opposite view to consumer-side fairness. A significant part of the research focuses on mitigating popularity bias, which occurs when popular items are given disproportional amounts of exposure by the recommender system. However, the broadness of the definition also covers research that is more similar to many consumer-side fairness approaches in considering fairness for groupings based on sensitive information from a provider perspective.

\subsubsection{Cold-start, Long-tail and Diversity}
Cold-start, long-tail, and diversity in recommender systems all make out similar concepts with partly overlapping causes and mitigation approaches: \textbf{Cold-start} specifically focuses on the scenario of providing good recommendations for new users or new items, through facing the challenge of comparatively little data for the new entity. \textbf{Long-tail} recommender system approaches more generally attempt to improve the recommendation of items that have few interactions compared to the more popular items or the analogous user-centric alternative in improving the recommendations for inactive users. Approaches that optimize for \textbf{Diversity} attempt to diversify the top-$k$ recommendations given to users, motivated by popularity bias issues and an effort to enhance the user experience. While the definitions differ in generality and perspective, they are sometimes used interchangeably in the literature, especially cold-start and long-tail. Provider-centric approaches of all three fields share similarities, or directly overlap, with popularity bias mitigation approaches proposed as provider-side fairness-focused recommenders. Similarly, user-centric approaches that seek to balance the performance of \textit{all individual users} may overlap with consumer-side fairness approaches and are included in this survey, given that they satisfy all acceptance criteria. The last point typically boils down to whether a fairness perspective is applied, i.e., posed as individual fairness, along with fairness evaluation. However, as a general strategy, research that seeks to balance the utility of user groups based on \textit{number of user interactions} have been excluded because they fit perfectly within the mature fields of cold-start or long-tail recommender systems, and are better represented when compared with such methods.

\subsubsection{IR Ranking Fairness}
IR Ranking is typically not personalized, i.e., the produced rankings are not affected by user-specific interaction with the system. Subsequently, IR Ranking fairness objectives usually have a provider-side point of view, e.g., balancing the exposure achieved by similar items or the representation of different item groups given non-personalized queries. The work of \cite{fair_rank_rec_2022} provides an overview of fairness in the IR Ranking setting.

\subsubsection{Group Recommendation}
Group recommendation approaches seek to recommend a set of items to a collection of users, e.g., recommending a travel destination for a group of friends with different preferences. This field frequently applies the term \textit{fairness} when explaining their motive of balancing the consideration of the various users in the groups, and most consumer-side fairness concepts covered in this survey are applicable within the groups or aggregated over groupings. However, because the field is specialized for a specific recommender system scenario and has received significant attention both before and after the notion of fairness gained traction, these approaches are not included in this survey.

\subsubsection{Privacy}
Privacy in recommender systems covers many approaches that seek to protect privacy in different stages of the recommender system pipeline. For instance, \textit{federated learning}  can be applied to mitigate the issues of having a centralized model that may be breached and mined for sensitive information \citep{federated_2020}. \textit{Differential privacy} has been applied to provide protection guarantees for the sensitive information of individual users \citep{differential_2010}. Some privacy approaches seek to completely remove the information of specific attributes within user representations or data, which overlaps with a class of fairness approaches that do the same with the intention of not having the attributes influence the recommendation. 

\subsection{Related Work} \label{sec:related_work}
There has been a recent surge in proposed surveys of fairness in recommender systems. \cite{fair_rank_rec_2022} surveys both fairness in IR ranking and recommender systems, while \cite{survey_deldjoo_2022,survey_wang_2022,survey_li_2022} focus on recommender systems. \cite{fair_rank_rec_2022} seeks to serve as an overview of fairness in both IR ranking and recommender systems, which makes it the broadest and most high-level survey among the ones considered relevant. They propose using multiple established concepts as a basis for their categorization, e.g., individual/group fairness and provider/consumer side fairness, as well as novel classifications of fairness incorporation methods. Because of the broad scope and since it is the oldest survey considered relevant, only two of the studies covered in this survey were covered in \cite{fair_rank_rec_2022}. 

\cite{survey_deldjoo_2022,survey_wang_2022,survey_li_2022} were all first made publicly available within a few months of each other in 2022 and all consider a broad scope comprising all types of fairness in recommender systems. This scope is wider than the one applied in this survey by covering provider-side fairness and group recommendation. Additionally, all three surveys also cover research that is theoretical in nature or performs analysis using established approaches and datasets, i.e., not necessarily proposing new models or methods. \cite{survey_deldjoo_2022} investigate the current state of fairness in recommender systems, and focus on charting how the available research is distributed with respect to different high-level concepts. They additionally propose a single-level categorization of the fairness metrics considered in the research they cover. \cite{survey_li_2022} has a more conceptual take and provides a thorough introduction to fairness theory and fairness in other machine learning fields before addressing their identified research through multiple different taxonomies based on binary concepts. Some of the fairness concepts they use to categorize the research are well established, like group/individual and consumer/provider, while others have not previously been focused on, like short-term/long-term and black box/explainable. \cite{survey_wang_2022} propose a hierarchical taxonomy of fairness definitions that are since used to categorize the optimization and fairness metrics applied in their identified studies.

Our work differs from previous surveys by specializing in tangible solutions proposed for consumer-side fairness in recommender systems. The specialization allows for a complete overview of the available literature and a higher focus on technical aspects to enhance comparisons. We also categorize our identified research using a new taxonomy centred on high-level notions of fairness interpretations and incorporation methods, which has a purposely high-level and general definition to be applicable and extendable to new fairness interpretations and incorporation methods. The completeness of our survey is exemplified by Table \ref{tab:overlap}, which indicates that when adjusting for time, the largest coverage overlap with the broader surveys only comprises 18 out of the 43 articles we identified in the same time interval.
\begin{table}[htb]
\tiny
\begin{center}
\caption{Table displaying the coverage in the most relevant surveys of the articles identified in this survey. Raw counts and percentages are presented, both adjusted and unadjusted for the publish date of the last considered article in each survey.}\label{tab:overlap}%
\begin{tabular}{l l l l l}
\toprule
 & \textbf{Adj. Coverage} & \textbf{\% Adj. Coverage} & \textbf{Tot. Coverage} &  \textbf{\% Tot. Coverage}\\
\cite{survey_deldjoo_2022} & 14/43 & 33\% & 14/47 & 30\% \\
\cite{survey_li_2022} & 18/43 & 42\% & 18/47 & 38\% \\
\cite{survey_wang_2022} & 16/41 & 39\% & 16/47 & 34\% \\
\bottomrule
\end{tabular}
\end{center}
\end{table}


\section{Methodology}
The methodology of this survey covers the systematic selection process applied for identifying and screening relevant studies, followed by the definition and justification of applied taxonomies, as well as descriptions of how the taxonomies are used to categorize and structure further discussion.

\subsection{Selection Process}
The selection process comprised the definition of concise acceptance criteria, identification of relevant publication indexes, query definition, two rounds of article screenings, and final in-depth reading of the remaining candidate articles. This section presents the acceptance criteria, details the queries and how they were defined, and presents a full overview of the number of studies involved in each step of the selection process. 

\subsubsection{Acceptance criteria}
Five acceptance criteria have been defined in line with our goals of examining the existing literature of tangible models considering consumer-side fairness in recommender systems:
\begin{enumerate}
    \item The study considers \textit{recommender systems}, see Section \ref{sec:rec_sys_def}.
    \item The study considers consumer-side fairness, either explicitly or through a multi-stakeholder focus.\\
    \textbf{Note}, Group recommendation and Long tail/Cold-start recommender systems are excluded, see Section \ref{sec:rel_rec_conc}.
    \item The study is published in a peer-reviewed conference or journal.
    \item The study proposes a novel model or method.
    \item The study evaluates the fairness of the proposed model or method.
\end{enumerate}

\subsubsection{Query Definition and Overview}
The keywords were kept general to avoid filtering out potential relevant research. The search queries were chronologically bound by 2016-01-01 and 2022-10-01, where the lower bound was set based on prior knowledge of the topic and preliminary querying for validation purposes. The topic started gaining noticeable traction in 2017, but the early adopters had three publications before this, \citep{kamishima_enhancement_2012,kamishima_efficiency_2013,kamishima_model-based_2016}. The first two articles do not appear to have inspired other researchers, but since 2016, there has been a gradual increase in the number of articles each year. 

The chronological bound was combined with the keyword combination “recommend*” and “fairness”, and both keywords had to be matched in the title, the abstract, or both. “Recommend” was given a wildcard suffix matcher to match both “recommender” and “recommendation”. A similar wildcard, “fair*”, was used instead of “fairness” in the DBLP index to compensate for not being able to match within the abstracts. Observations in known research and research found through preliminary querying confirmed that all articles that matched “fair” in the title also matched “fairness” when considering the abstract. The wildcard was only used in title-only queries since it significantly increased the number of false positives when matching in both title and abstract. Fairness is becoming a well-established concept within the ML community, and most, if not all, research uses the full term at least once before potentially switching over to the shorthand “fair”. 

\begin{figure}[!htb]
\centering
\caption{A PRISMA flow diagram illustrating the full selection process.}
\includegraphics[width=0.8\textwidth]{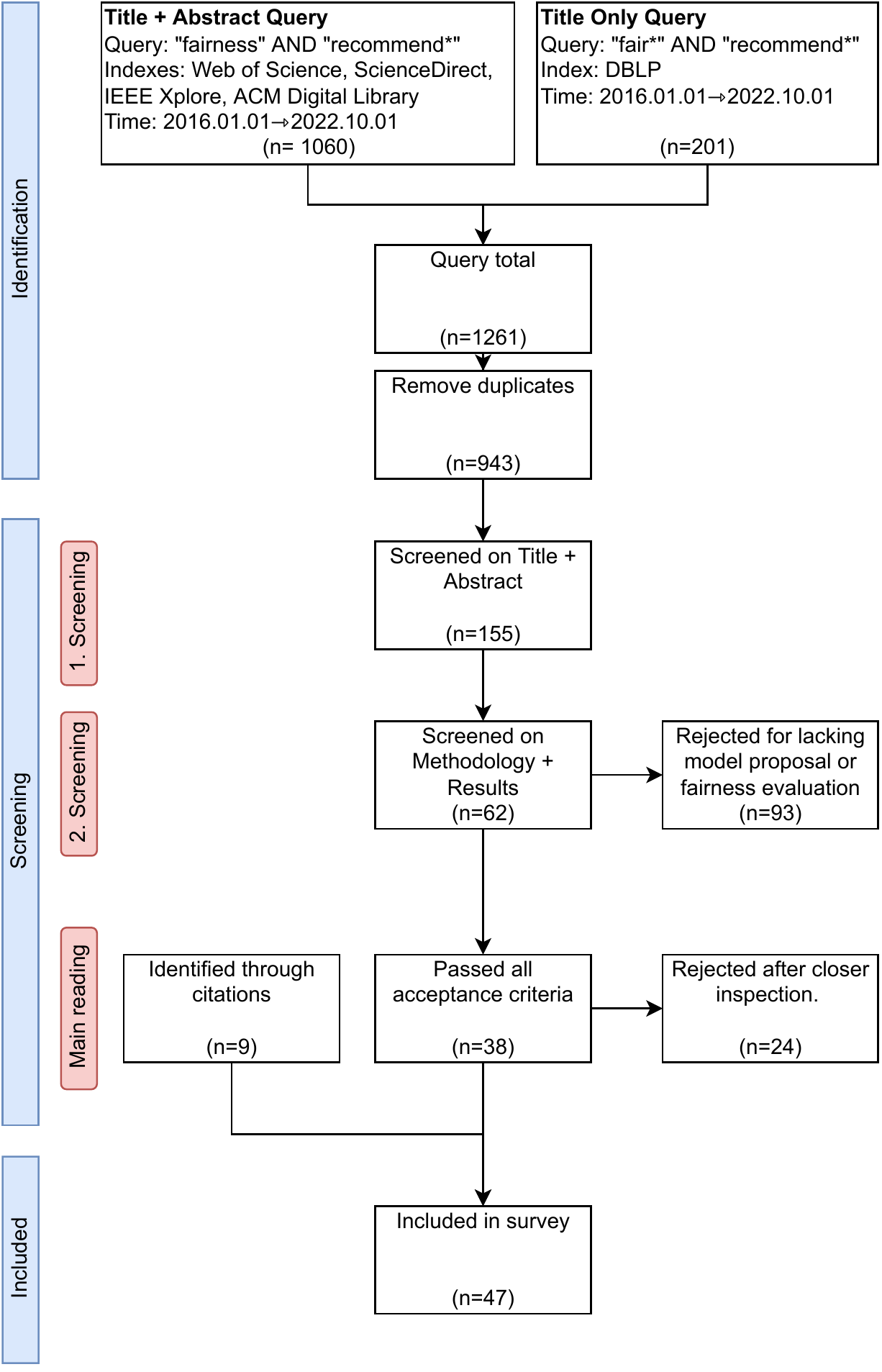}
\label{fig:selection}
\end{figure}

The full selection process is detailed in a PRISMA flow diagram \citep{prisma_2021} in Figure \ref{fig:selection}.

\subsection{Taxonomy}
While there have been previous attempts at proposing novel taxonomies for categorizing fairness approaches in recommender systems based on which Fairness Interpretation is pursued, we argue that there are alternative taxonomies that offer additional insight and value. The most recent taxonomy is proposed by \cite{survey_wang_2022}, who first proposes splitting between process and outcome focus, then two alternatives for splitting outcome-focused fairness on target and concept. One challenge when applying this to consumer-side fairness research is that many of the named \textit{concept-based} fairness categories do not occur that often, and the vast majority of identified research would be classified as either optimizing and evaluating for \textit{Process Fairness} or \textit{Consistency Fairness}. We also argue that there may be value in further separating different high-level Fairness Interpretations, e.g., \textit{Consistency Fairness} may consider distance notions that only compare the distribution of predictions given to different groups, but it can also consider distance notions that measure differences in how the predictions of the same groups match the targets.

We propose a new taxonomy centred on which high-level Fairness Interpretation is considered when optimizing and evaluating models. Besides resulting in a balanced segmentation of the identified research, the taxonomy separates key differences in mentality when approaching fairness, some of which fundamentally conflict with each other. To further structure and analyze the research, we propose applying two other, more established, concepts which detail how/when the fairness consideration is incorporated and which type of recommender model is applied, respectively.


\subsubsection{Fairness Interpretation Taxonomy}\label{sec:fair_cat}
\begin{figure}[htb]
\centering
\caption{The proposed taxonomy based on Fairness Interpretation.}
\includegraphics[width=0.8\textwidth]{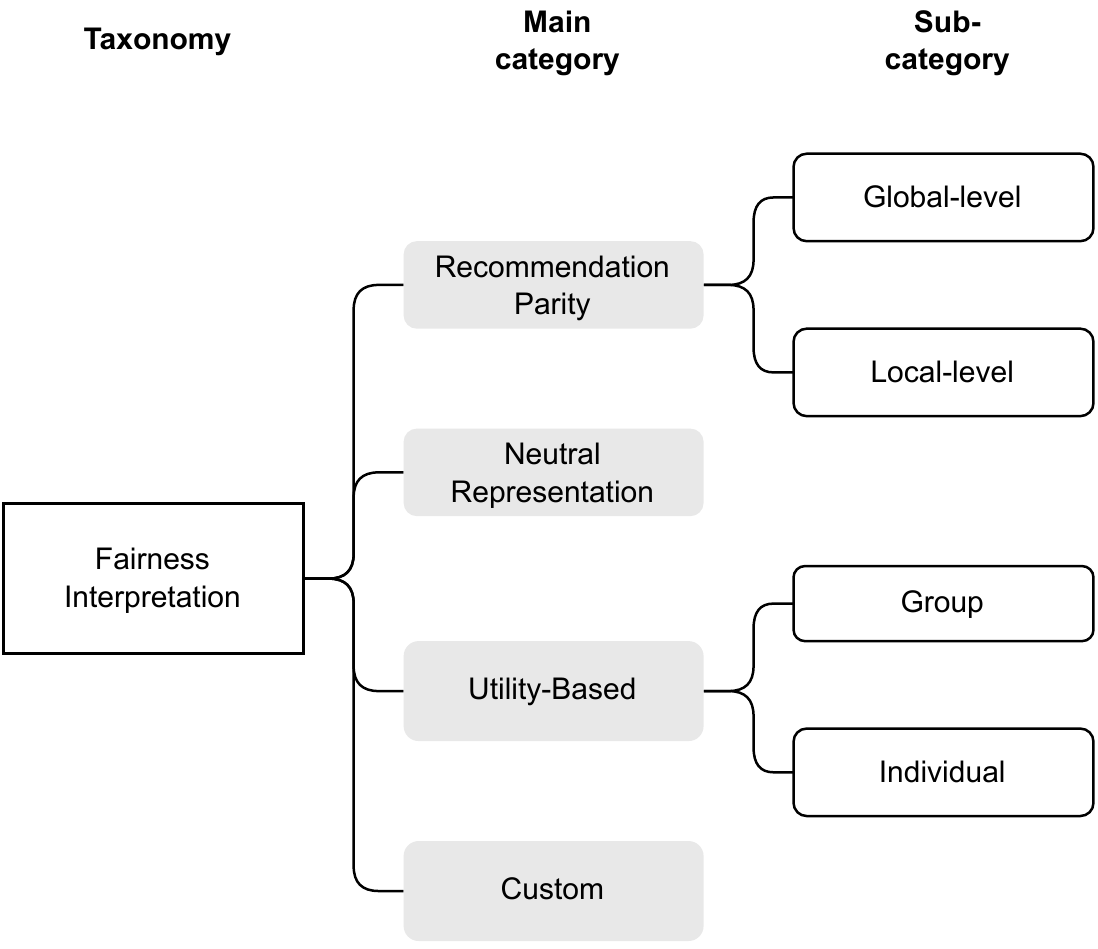}
\label{fig:inter}
\end{figure}

\begin{figure}[htb]
\centering
\caption{Diagram that illustrates the high-level differences between the three non-Custom Fairness Interpretations in a scenario where the sensitive groups $s_1$ and $s_2$ display different preferences and the base recommender perform better for $s_1$. The preferences and recommendations given to the groups are illustrated as probability distributions, while model representations are projected into two-dimensional scatterplots. The Recommendation Parity interpretation idealizes when the recommendation distributions overlap, while a Utility-Based interpretation requires that the respective recommendation distributions match and mismatch the ``true" distributions equally. The Neutral Representation interpretation is optimized to move from the case where representations of different groups can be separated into distinct clusters to the case where the clusters overlap or are indistinguishable.
}
\includegraphics[width=0.60\textwidth]{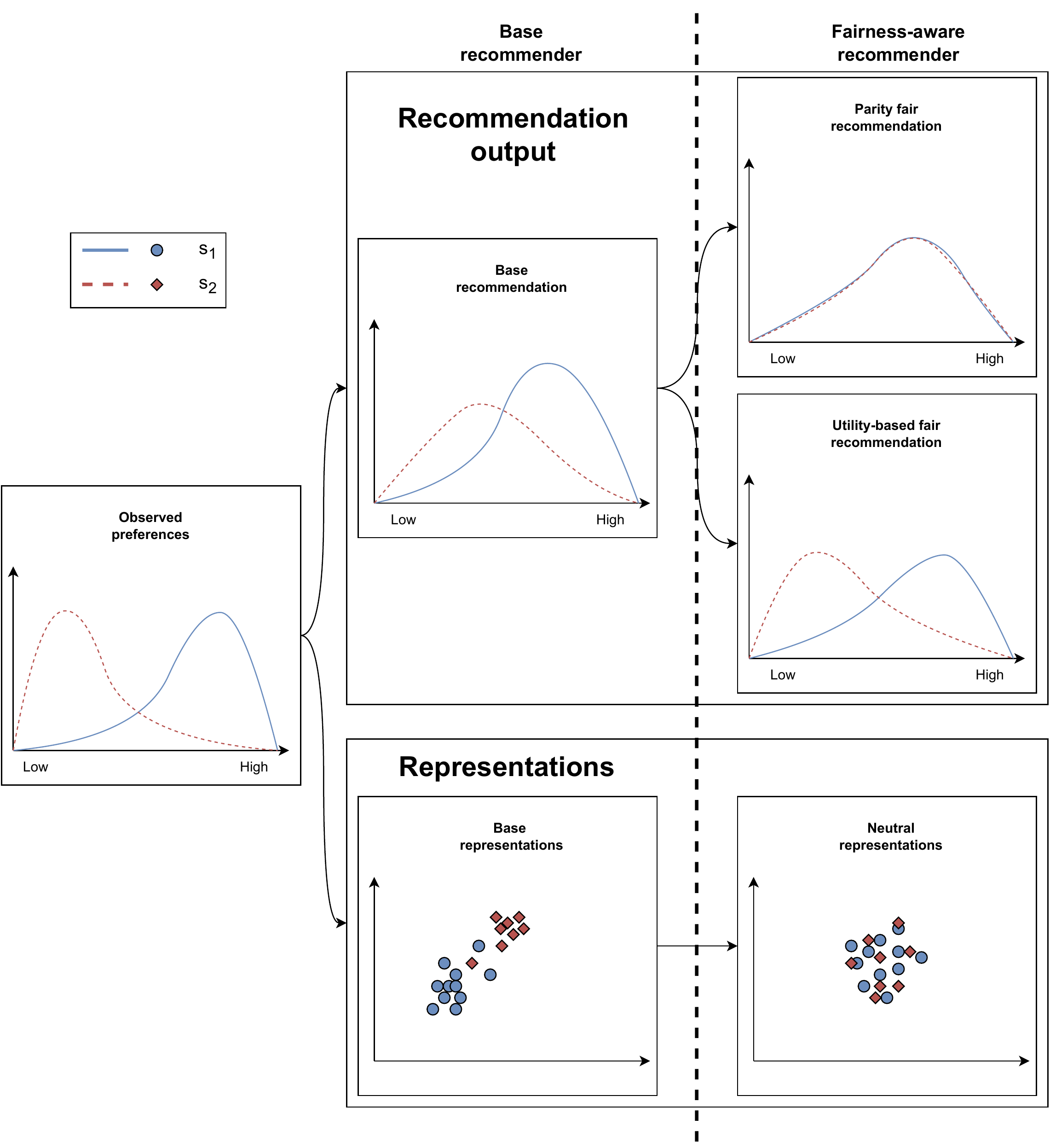}
\label{fig:inter_diag}
\end{figure}

While several fairness definitions from the fields of law and psychology have been formally defined for machine learning, see Section \ref{sec:formal}, they cannot trivially be applied for categorizing the studies considered in this survey. 
The formal definitions are occasionally implemented as metrics, but since they mostly consider the model's outcome, it is challenging to define how they should be adhered to during optimization. Another challenge is that some of these definitions are conceptually similar and only differ in minute details. We instead propose categorizing the Fairness Interpretation on a higher and more conceptual level, while remaining compatible with the more low-level formal definitions. For instance, Equality of Opportunity and Equalized Odds share a high-level objective of balancing utility measures evaluated for different sensitive groups that consider both the predictions and how they match the targets. Two identified interpretations have further been assigned sub-categories for finer distinctions between similar concepts. The full taxonomy is illustrated in Figure \ref{fig:inter}, and the different interpretations are further described in the following sections and illustrated in Figure \ref{fig:inter_diag}.

\paragraph{Recommendation Parity}
Recommendation Parity methods consider the distribution of ratings, rankings or preference scores. The Recommendation Parity-based approaches and metrics are strictly applied for group fairness views and are optimized when the recommendation distribution given to different users is similar. The strong focus on recommendation distribution, while completely disregarding differences in the achieved utility of different sensitive groups, makes Recommendation Parity a highly contrasting Fairness Interpretation to Utility-Based fairness and various Custom Fairness Interpretations.

Recommendation Parity for consumer side fairness can be further split based on at which level the parity is optimized or measured. Some optimize and measure parity at a \textbf{global} level, while others consider parity in the rating of individual \textbf{items} or \textbf{item groups}. The former is less constricting, as sensitive groups may disagree on individual items as long as the disagreements cancel out globally, i.e., if one group is more fond of an item than another, perfect global parity is regained if an identical reversed profile exists for another item. Local-level parity requires different sensitive groups also to rate/prefer individual items/item groups similarly. 

\paragraph{Neutral Representation}
Adversarial and orthogonality approaches both consider the case where the model is oblivious to the sensitive attributes of the users \textit{fair} and achieve this by making sure the intermediate representations of users do not reflect any of their sensitive information. A special case of this, achievable through modelling latent variables independent of the sensitive variables in causal models, is defined as Counterfactual Fairness by \cite{counterfactual_2017}. Counterfactual Fairness is considered an individual form of fairness, in the sense that changing the sensitive attributes in individual cases should not affect the outcome. In the more general case, the indiscriminate and global removal of sensitive information based on correlations between sensitive attributes and representations arguably resembles Recommendation Parity more, as recommender systems using sensitive-neutral representations should inherently lead to sensitive-neutral recommendations. However, given the unique perspective of focusing on representations rather than the recommendations, characteristic optimizations, representation-centric evaluation, and overall prevalence of approaches, a dedicated interpretation category for neutral user representations is still deemed warranted.

There are no conceptual sub-categories of this Fairness Interpretation, but four main strategies for neutralizing representations have been identified: \textit{Adversarial}, \textit{Orthogonality}, \textit{Fairness-aware sampling} and \textit{Probabilistic}. Succinctly, adversarial approaches train classification models to discriminate sensitive attributes from user representations. These models are trained in parallel with the main model(s), and their insights are used to inform the main model how it should be updated to make it harder for the adversarial to discern sensitive attributes accurately. Proposition 2 from \cite{gan_2014} propose that given a generator model and an adversarial model with sufficient capacity, the generator can be trained to generate data adhering to the original data distribution. By replacing the considered task with the task of distinguishing a binary sensitive attribute given representations, a similar proposition can be made for adversarial approaches for producing neutral representations, i.e., given sufficient capacities, neutral representations are achievable.

Orthogonality approaches utilize the notion of representation space in considering the representations to be vectors in a high-level semantic space. They identify sensitive dimensions in said space and apply different methods to ensure that the representations are orthogonal to these dimensions. Thus, the representations will ideally not contain any intrinsic sensitive information themselves.

Fairness-aware sampling and probabilistic modelling have also been used to reduce the amount of sensitive information in representations of different flavours. Fairness-aware sampling approaches alter sampling done when training representations to be more diverse with respect to sensitive attributes. Probabilistic approaches for producing neutral representations typically explicitly provide or model sensitive information during training,  which disincentivizes the rest of the model from modelling the same sensitive information. 

\paragraph{Utility-Based Fairness}
Utility-Based fairness is a comprehensive class focusing on differences in utility aspects of the recommendations given to different sensitive \textit{groups} or different \textit{individuals}. That is, the distribution of recommendations of different individuals/groups is free to differ significantly as long as it does not affect the utility of various entities in a way deemed unfair relative to the utility of other entities. The one key requirement of the utility that separates this interpretation from the other interpretations is that the utility functions considered \textbf{must be tied to the ground truth targets} in some way. That is, correctly recommending an item will directly affect the utility value. 

Numerous variations of optimization terms and metrics fall under this interpretation, as the notion of fairness may depend on the utility measure and the scenario. The observed variations are based on equal recommendation metric scores, equal utility variance, or the loss of individual utility in two-sided fairness settings etc.

\paragraph{Custom}
The final interpretation encompasses measures of adherence to custom fairness definitions and is a collection of optimizations that do not fit in any other interpretation category, e.g., parity with respect to derived attributes or balancing of custom utility measures that are independent of the ground-truth targets.

\subsubsection{Fairness Incorporation}
\begin{figure}[htb]
\centering
\caption{Fairness Incorporation categories.}
\includegraphics[width=0.8\textwidth]{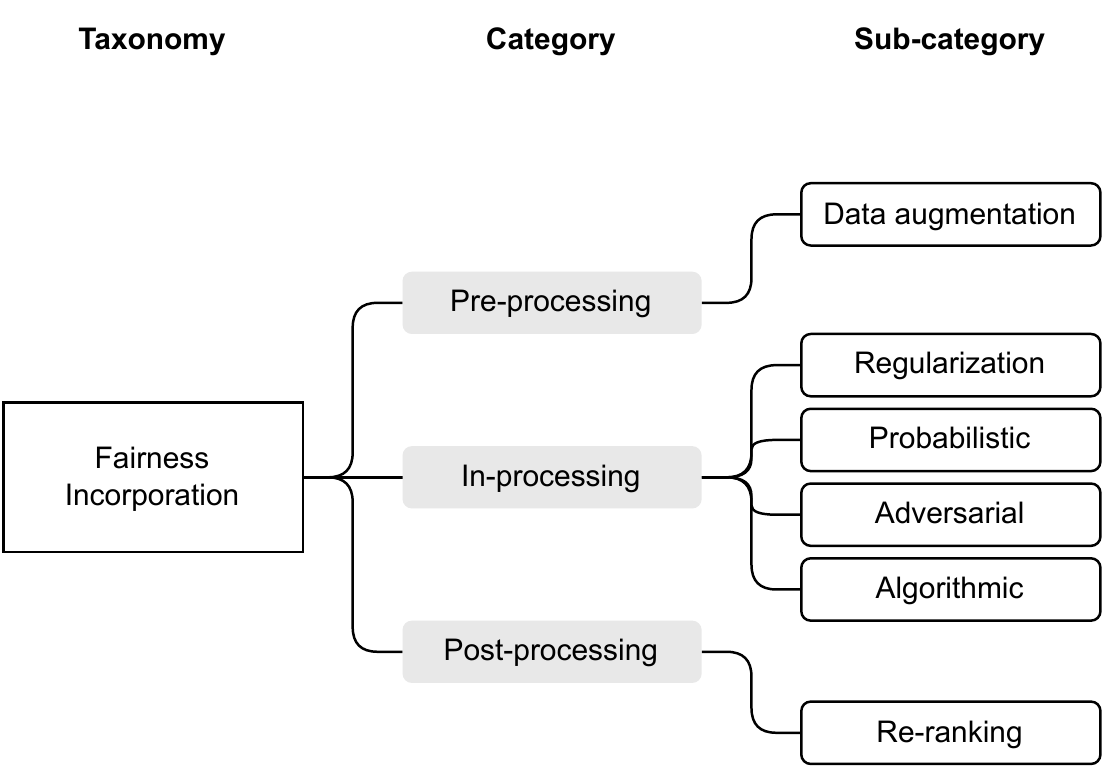}
\label{fig:incorp}
\end{figure}

The notions of pre-, in- and post-processing methods for injecting additional specialized considerations in approaches are well established both within the field of recommender system fairness and machine learning as a whole \citep{fair_ml_survey_2020,bias_ml_2021,survey_deldjoo_2022}, and categorize if the methods take place before, within or after the application of the model they enhance. To structure the identified research, we propose extending this taxonomy with an additional level to represent better the observed variety of approaches applied to incorporate fairness awareness. The second level contains a single sub-category each for pre- and post-processing approaches, but we propose four sub-categories to cover the diversity of in-processing methods. The full overview is illustrated in Figure \ref{fig:incorp}, and each proposed sub-category has been given a brief description in this section. 

\paragraph{Data Augmentation:}
The only sub-category of pre-processing methods is Data Augmentation, which covers all methods that inject fairness consideration by augmenting the model's input data.

\paragraph{Loss enhancement:}
Loss enhancement methods encourage fairness consideration through additional terms in the loss used for optimizing the model. Positive aspects of loss enhancement methods are that they can be applied to many model types, are flexible in definition, and can significantly change predictions through minimal changes to an approach. However, extra loss terms do not inherently improve the modelling capacity of a model but may introduce more complex dynamics that would benefit from more modelling capacity or changes to the model architecture.

\paragraph{Probabilistic:}
The probabilistic fairness approaches apply probabilistic concepts to encourage independence of recommendation and sensitive features, apply soft constraints, or filter out sensitive information. Unlike the other in-processing sub-categories, probabilistic fairness approaches are not easily achieved through a smaller extension to an arbitrary model. This variation of Fairness Incorporation usually requires that the applied model is probabilistic in nature itself, at least partially.

\paragraph{Algorithmic:}
Algorithmic approaches incorporate fairness by changing smaller aspects of an existing algorithm or through one-time processing, e.g., through selective sampling or removal of sensitive projections in representation space.

\paragraph{Adversarial:}
Adversarial approaches apply adversarial models to identify sensitive information from intermediate representations, to identify how the main model can be updated to better filter out sensitive information.

\paragraph{Re-Ranking:}
Re-ranking approaches re-rank the recommendation of one or more base recommender systems according to new or changed objectives, e.g., introducing fairness objectives that are optimized along with regular recommendation utility.

\subsubsection{Model Types}
A third categorization system is used to categorize approaches by which type of recommender system model architectures they fall under. The model type can affect how fairness awareness can be incorporated and influence the general recommendation task. Comparing approaches based on the same model types is also enhanced by sharing similar implementation details and premises. Several model groups have been defined based on the prevalence and shared concepts and can be found listed by an acronym and a description in the following list.
\begin{itemize}
    \item \textbf{CF: } Neighbourhood-based Collaborative filtering.
    \item \textbf{MF: } Matrix-Factorization.
    \item \textbf{NCF: } Neural Collaborative Filtering, taken to mean neural network-based collaborative filtering methods that more specialized model groups do not cover.
    \item \textbf{Graph: } Various Graph-based models and methods. Graph Neural Networks, Graph Convolutional Networks, Graph Embeddings etc.
    \item \textbf{AE: } (Variational) Auto Encoders.
    \item \textbf{Probabilistic: } Various Probabilistic models and methods. Probabilistic Soft Logic, Latent models, Bayesian Networks etc.
    \item \textbf{Classification: } Various Classification methods. Random Forest, Gradient Boosting, Naive Bayes etc.
    \item \textbf{Bandit: } Contextual Bandit.
\end{itemize}

\subsubsection{Structuring of Main Discussion}
The three different categorizations will all be used when discussing and comparing the identified approaches. Three sections are reserved for pre-, in- and post-processing Fairness Incorporation approaches, and their content is structured by the corresponding Fairness Incorporation sub-categories and both Fairness Interpretation and model types. The Fairness Interpretation taxonomy is used in an overview and for a focused comparison of fairness optimization. In contrast, model types are used to structure a more general technical discussion to highlight comparable implementational choices.

\subsection{Full Model Overview}
This section presents a preliminary analysis and overview of all identified research according to model type and Fairness Incorporation method. The motive is to put the topic into the broader context of general recommender systems and to provide an overview of all covered research. A full overview is found in Table \ref{tab:tax}. Note that the same article may fall under multiple types of Fairness Incorporation and model types, since the proposed approach may apply multiple types of Fairness Incorporation strategies and be applied on multiple base models. Also note that while many incorporation methods can be applied to multiple different model types, especially pre- and post-processing methods, only observed combinations are covered. The fact that a method is adaptable for other model types does not guarantee that un-documented combinations will achieve similar results or improvements. Furthermore, the current trends of the field are better reflected when keeping to the combinations that have been actively researched.
\begin{sidewaystable}
\tiny
\sidewaystablefn%
\begin{center}
\begin{minipage}{\textheight}
\caption{All covered research categorized by Fairness Incorporation method and recommender system model type.}\label{tab3}
\begin{tabular*}{\textheight}{p{0.07\textwidth}p{0.13\textwidth}p{0.13\textwidth}p{0.13\textwidth}p{0.13\textwidth}p{0.13\textwidth}p{0.13\textwidth}}
\toprule%
& \textbf{Pre-processing} & \multicolumn{4}{@{}c@{}}{\textbf{In-processing}} & \textbf{Post-processing} \\\cmidrule{3-6}%
 & \textbf{Data augmentation} & \textbf{Loss Enhancement} & \textbf{Probabilistic} & \textbf{Algorithmic} & \textbf{Adversarial} & \textbf{Re-ranking} \\
\midrule
\textbf{CF} & \cite{slokom_towards_2021} & \cite{burke_balanced_2018} &  &  &  & \cite{patro_incremental_2020},\newline\cite{ashokan_fairness_2021},\newline\cite{wu_tfrom_2021} \\
& & & & & & \\
\textbf{MF} & \cite{rastegarpanah_fighting_2019},\newline\cite{fang_fairroad_2022},\newline\cite{slokom_towards_2021} & \cite{kamishima_enhancement_2012},\newline\cite{kamishima_efficiency_2013},\newline\cite{kamishima_considerations_2017},\newline\cite{yao_beyond_2017},\newline\cite{kamishima_recommendation_2018},\newline\cite{zheng_fairness_2018},\newline\cite{wan_addressing_2020},\newline\cite{yao_personalized_2021} &  &  & \cite{resheff_privacy_2019},\newline\cite{li_towards_2021},\newline\cite{wu_learning_2021} & \cite{edizel_fairecsys_2020},\newline\cite{patro_fairrec_2020},\newline\cite{patro_incremental_2020},\newline\cite{wu_tfrom_2021},\newline\cite{ashokan_fairness_2021},\newline\cite{biswas_toward_2021},\newline\cite{lorenz_two_2021},\newline\cite{wu_multi-objective_2022} \\
& & & & & & \\
\textbf{NCF} &  & \cite{bobadilla_deepfair_2021},\newline\cite{islam_debiasing_2021},\newline\cite{wu_fairness-aware_2021},\newline\cite{li_leave_2021} &  & \cite{islam_mitigating_2019},\newline\cite{islam_debiasing_2021},\newline\cite{li_fairsr_2022} & \cite{li_towards_2021},\newline\cite{wu_fairness-aware_2021}\newline\cite{wu_selective_2022},\newline\cite{wei_comprehensive_2022},\newline\cite{closing_2022} & \cite{wu_multi-objective_2022}\\
& & & & & & \\
\textbf{Graph} &  & \cite{liu_dual_2022},\newline\cite{liu_self-supervised_2022} & \cite{buyl_debayes_2020},\newline\cite{li_toward_2022} & \cite{rahman_fairwalk_2019},\newline\cite{xu_fair_2021},\newline\cite{li_fairsr_2022} & \cite{bose_compositional_2019},\newline\cite{wu_learning_2021},\newline\cite{xu_fair_2021},\newline\cite{liu_dual_2022},\newline\cite{liu_mitigating_2022},\newline\cite{liu_self-supervised_2022} & \\
& & & & & & \\
\textbf{Probabilistic} &  &  & \cite{kamishima_model-based_2016},\newline\cite{farnadi_fairness-aware_2018},\newline\cite{buyl_debayes_2020},\newline\cite{dickens_hyperfair_2020},\newline\cite{frisch_co-clustering_2021},\newline\cite{li_toward_2022} &  &  & \cite{dickens_hyperfair_2020}\\
& & & & & & \\
\textbf{AE} &  & \cite{li_leave_2021},\newline\cite{borges_f2vae_2022} &  &  & \cite{borges_f2vae_2022} & \\
& & & & & & \\
\textbf{Classification} &  &  &  &  &  & \cite{paraschakis_matchmaking_2020}\\
& & & & & & \\
\textbf{Bandit} &  & \cite{huang_fairness-aware_2021} &  &  &  &\\
\bottomrule
\label{tab:tax}
\end{tabular*}
\end{minipage}
\end{center}
\end{sidewaystable}

\subsubsection{Model Analysis}
Some clear trends can be observed in the full table. The field has experienced rapid growth, with most research taking place in the most recent years. Pure loss enhancement approaches saw a lot of attention among the early adopters, especially when used together with matrix factorization models but are currently rarely used as the sole Fairness Incorporation method. Re-ranking methods saw a similar burst of attention in 2020 and 2021. Still, they did not dominate the field, unlike early loss enhancement approaches, as multiple other directions were researched simultaneously. Adversarial approaches were slow to appear but have since become popular while seemingly being used with a more varied selection of base recommender system types. Bayesian and algorithmic approaches are the smallest in-processing groups but are characterized by being pretty evenly distributed across time and being applied with specific types of recommender systems. There also appears to be a recent trend of applying multiple Fairness Incorporation strategies instead of relying solely on a single strategy. 

\section{Pre-Processing Methods}
While numerous studies consider the effect of data augmentation, we  only found three papers that pass all acceptance criteria. In particular, several candidates were rejected for not proposing formalized approaches or not presenting an evaluation of the achieved fairness. Pre-processing methods comprise the smallest Fairness Incorporation main category.

\subsection{Fairness Optimization}
\begin{table}[htb]
\begin{center}
\caption{Overview of the identified pre-processing approaches structured by the Fairness Interpretation and Fairness Incorporation of their optimization. Approaches that consider multiple Fairness Interpretations are listed in multiple rows.}\label{tab:fair_pre}%
\begin{tabular}{p{0.27\textwidth}p{0.2\textwidth}p{0.4\textwidth}}
\toprule
& & \textbf{Data Augmentation}\\
\midrule
\textbf{Recommendation Parity} & \textbf{Global} & \cite{fang_fairroad_2022}\\
& & \\
& \textbf{Local} & \\
\midrule
\textbf{Neutral\newline Representation} & & \\
\midrule
\textbf{Utility-Based} & \textbf{Group} & \cite{rastegarpanah_fighting_2019}\newline\cite{fang_fairroad_2022}\\
& & \\
& \textbf{Individual} & \cite{rastegarpanah_fighting_2019}\\
\midrule
\textbf{Custom} & & \cite{slokom_towards_2021}\\
\bottomrule
\end{tabular}
\end{center}
\end{table}

\subsubsection{Utility-Based Fairness}
\cite{rastegarpanah_fighting_2019,fang_fairroad_2022} propose exploiting collaborative filtering dynamics by training new synthetic users that will influence the recommendation of the real users. When training synthetic user, \cite{rastegarpanah_fighting_2019} enhances the loss by adding terms for  penalizing both the group-level and the individual-level variance of mean squared errors. In contrast, \cite{fang_fairroad_2022} utilize similar loss terms based on the utility metrics proposed by \cite{yao_beyond_2017}, see Section \ref{sec:beyond}, and also global Recommendation Parity.

\subsubsection{Custom}
The fairness optimization proposed by \cite{slokom_towards_2021} shares similarities with in-processing approaches optimizing for Neutral Representations but alters the input data to remove correlation between the user profiles and the sensitive attributes of the users, instead of altering the intermediate user representations. The approach achieves this by adding items that are indicative of the other sensitive groups, identified using auxiliary models, to the user profiles, and they also explore removing items at random or based on how indicative they are of the actual sensitive attributes of the user.

\subsection{Architecture and Method}
The three selected papers all propose pre-processing methods that can be applied to a wide variety of model types, but all have used matrix factorization as one of their base recommender models. 
\cite{rastegarpanah_fighting_2019} propose a method for training supplementary data that can influence polarization and improve individual and group fairness. 
The key insight is that 
introducing additional users will affect the recommendation of the original users. This insight is exploited by adding a few synthetic users with accompanying data and allowing gradients from loss terms designed to influence the polarization and fairness to flow back to these. 
Further, they propose two computationally cheap  heuristics. 
The fairness optimization idealizes equal utility of individual users and groups explicitly. \cite{fang_fairroad_2022} apply the same base approach but focus on optimizing multiple fairness objectives more efficiently and smoothly by projecting the gradients of different objectives onto each other if they conflict. 
The fairness objectives fall under both Utility-Based fairness and Recommendation Parity.

\cite{slokom_towards_2021} modify the data of existing users through an extension of the approach proposed by \cite{weinsberg_blurme_2012} instead of training new ones. 
An auxiliary logistic regression model is trained to tell how indicative items are of the gender of the users that like them. This information is used to select items to be added or removed from user data to make the data less indicative of gender. 
The addition process specifically intersects lists of indicative items with recommendations from a user-based collaborative filtering model to motivate the addition of relevant items.

\section{In-Processing Methods}
In-processing methods are the most represented among the main categories, and their dominance has been constant since the birth of the field. They are characterized by being the most specialized approaches, as the base models themselves are adapted and changed.

\subsection{Fairness Optimization}
\begin{sidewaystable}
\tiny
\sidewaystablefn%
\begin{center}
\begin{minipage}{\textheight}
\caption{Overview of the identified in-processing approaches structured by the Fairness Interpretation and Fairness Incorporation of their optimization. Approaches that consider multiple Fairness Interpretations and Fairness Incorporation methods are listed in multiple rows and columns.}\label{tab:fair_in}
\begin{tabular*}{\textheight}{p{0.13\textwidth}p{0.07\textwidth}p{0.175\textwidth}p{0.175\textwidth}p{0.175\textwidth}p{0.175\textwidth}}
\toprule%
 & & \textbf{Loss Enhancement} & \textbf{Probabilistic} & \textbf{Algorithmic} & \textbf{Adversarial} \\
\midrule
\textbf{Recommendation Parity} & \textbf{Global} & \cite{kamishima_enhancement_2012}\newline\cite{kamishima_efficiency_2013}\newline\cite{kamishima_considerations_2017}\newline\cite{kamishima_recommendation_2018} & \cite{kamishima_model-based_2016}\newline\cite{dickens_hyperfair_2020} & & \\
& & & & & \\
& \textbf{Local} & \cite{kamishima_efficiency_2013}\newline\cite{islam_debiasing_2021} & \cite{farnadi_fairness-aware_2018} & & \\
\midrule
\textbf{Neutral Representation} & & \cite{wu_fairness-aware_2021}\newline\cite{liu_dual_2022}\newline\cite{liu_self-supervised_2022} & \cite{buyl_debayes_2020}\newline\cite{frisch_co-clustering_2021}\newline\cite{li_toward_2022} & \cite{rahman_fairwalk_2019}\newline\cite{islam_mitigating_2019}\newline\cite{islam_debiasing_2021}\newline\cite{xu_fair_2021}\newline\cite{li_fairsr_2022} & \cite{resheff_privacy_2019}\newline\cite{wu_fairness-aware_2021}\newline\cite{xu_fair_2021}\newline\cite{borges_f2vae_2022}\newline\cite{bose_compositional_2019}\newline\cite{li_towards_2021}\newline\cite{wu_learning_2021}\newline\cite{liu_dual_2022}\newline\cite{wu_selective_2022}\newline\cite{liu_mitigating_2022}\newline\cite{liu_self-supervised_2022}\newline\cite{wei_comprehensive_2022}\newline\cite{closing_2022}\\
\midrule
\textbf{Utility-Based} & \textbf{Group} & \cite{yao_beyond_2017}\newline\cite{zheng_fairness_2018}\newline\cite{wan_addressing_2020}\newline\cite{huang_fairness-aware_2021}\newline\cite{yao_personalized_2021}\newline\cite{liu_dual_2022}\newline\cite{borges_f2vae_2022} & \cite{dickens_hyperfair_2020} & \cite{li_leave_2021} & \\
& & & & & \\
& \textbf{Individual} & & & & \\
\midrule
\textbf{Custom} & & \cite{burke_balanced_2018}\newline\cite{wan_addressing_2020}\newline\cite{bobadilla_deepfair_2021} & & & \\
\bottomrule
\label{tab:in_fair}
\end{tabular*}
\end{minipage}
\end{center}
\end{sidewaystable}

\subsubsection{Recommendation Parity}\label{sec:inproc:fairness:recparity}

Optimization of Recommendation Parity fairness, i.e.,  the statistical parity of recommendations given to different sensitive groups, is mainly found among the in-processing methods and was popular during the field's early years. 

\paragraph{Global Recommendation Parity:}
\cite{kamishima_efficiency_2013,kamishima_considerations_2017} propose adding loss terms for matching mean rating and preference of different sensitive groups, while \cite{dickens_hyperfair_2020} devise a probabilistic soft logic rule of similar design for the same goal. 

More comprehensive approaches for matching global recommendation distributions beyond the first momentum are proposed by \cite{kamishima_enhancement_2012,kamishima_model-based_2016,kamishima_recommendation_2018}.
\cite{kamishima_enhancement_2012,kamishima_recommendation_2018} introduce different loss terms for minimizing the mutual information of the ratings and the sensitive groups in matrix factorization.
In a slightly different approach, \cite{kamishima_model-based_2016} apply a latent factor model where the rating variable is considered independent of the sensitive group variable and optimizes their model using the Expectation Maximization algorithm.

\paragraph{Local Recommendation Parity:}
In the case of local Recommendation Parity, all relevant research we have found only considers the first moment when matching the recommendations of different sensitive groups. 
\cite{kamishima_efficiency_2013} propose adding a loss term that penalizes the squared difference of item ratings between different sensitive groups as an alternative to the already mentioned global version. 
Similarly, \cite{islam_debiasing_2021} apply the same idea but opt for an absolute difference instead of a squared difference. The probabilistic soft logic approach proposed in \cite{farnadi_fairness-aware_2018} defines rules for encouraging both item-group and item-level parity.

\subsubsection{Neutral Representation}
The objective of Neutral Representation fairness cannot be achieved without altering the model, thus, it is only pursued by in-processing approaches. 
Optimization of this Fairness Interpretation can be achieved by applying different strategies for filtering out intrinsic sensitive information in representations within the model. 
The following paragraphs are structured by the technique applied to achieve neutral representations, see also Fig \ref{fig:incorp}.

\paragraph{Adversarial:}
The approaches proposed by \cite{resheff_privacy_2019,wu_fairness-aware_2021,xu_fair_2021,borges_f2vae_2022,closing_2022} all apply adversarial models directly on model representations. 
\cite{resheff_privacy_2019} pass the latent user factors of their matrix factorization approach to their adversarial, while \cite{wu_fairness-aware_2021} do the same with one of the multiple user representations they train in a composite NCF model. \cite{xu_fair_2021} feed their adversarial model a linear combination of the user representation in a base recommender model and a representation they base on an auxiliary knowledge graph for modelling sensitive user attributes. \cite{closing_2022} propose a neural classification model and applies an adversarial model on a hidden layer in the said model. Finally, \cite{borges_f2vae_2022} apply an adversarial model to discriminate the latent representation in their variational autoencoder-based model.

A slightly more intricate scheme is proposed by \cite{wei_comprehensive_2022}, who concatenate the observed ratings to the representations that are fed to the adversarial model, which they argue will improve the neutrality of the representation and also make the representations independent with respect to the sensitive attribute conditioned on the observed ratings. They further add a second adversarial model, which is fed predicted ratings along with corresponding observed values and item embeddings.

\cite{bose_compositional_2019,li_towards_2021} argue for letting users dynamically decide which sensitive attributes they are comfortable with the model using. To support this, both propose training optional filters for filtering out different types or combinations of sensitive information from user representations in graph- and matrix-factorization models. The filters are trained using adversarial models. 
A similar approach is proposed by \cite{wu_selective_2022}, who train \textit{adaptors} \citep{adaptor_2019} within the \textit{transformers} \citep{transformer_2017} that make out their model. The adaptors dynamically filter out different combinations of sensitive attributes based on user- and task-based settings in a sequential recommendation setting.

\cite{wu_learning_2021,liu_self-supervised_2022,liu_mitigating_2022,liu_dual_2022} all consider graph neural network methods and the construction of higher-order graph representations by accumulating neighbouring representations in the recommendation graph. The approaches apply adversarial models to discourage the encoding of sensitive attributes in the user- and item-level representations, which also mitigate the accumulation of sensitive information in the higher-order neighbourhood representations.  \cite{liu_dual_2022} further supplements the adversarial discrimination loss with a loss term on the covariance of the predicted attribute and the actual sensitive attribute. \cite{liu_self-supervised_2022} instead designs and utilizes self-supervising loss terms to enhance the representations and mitigate imbalance issues caused by imbalanced sensitive attributes.

\paragraph{Orthogonality:}
Orthogonality-based approaches apply additional loss terms or explicit removal of sensitive projections to make representations orthogonal to explicit or implicit sensitive dimensions in the representation space. \cite{wu_fairness-aware_2021} model two separate user representations: one for inferring sensitive information and one for providing neutral representations. They devise a loss term that encourages the two representations to be orthogonal and further encourages the neutrality of the second representation through an adversarial approach.

A more explicit approach is pursued by \cite{islam_mitigating_2019,islam_debiasing_2021}, where a post hoc step infers sensitive dimensions in the representation space by taking the difference of the mean representation of each sensitive group. The projections of the sensitive dimension onto each representation are then explicitly subtracted. In the case of \cite{islam_debiasing_2021}, the orthogonality processing supplements the Recommendation Parity loss term (see Section~\ref{sec:inproc:fairness:recparity}).

\paragraph{Sampling Based Representation Training:}
\cite{rahman_fairwalk_2019,li_fairsr_2022} both adjust the sampling strategy used when training representations. 
\cite{rahman_fairwalk_2019} proposes to balance the sampling of the next user according to sensitive groups when training graph representations using random walks. In contrast, \cite{li_fairsr_2022} adjust the probability of sampling the triplets needed for training knowledge graph representations in a manner that balances out the correlation of sensitive groups and items across all users.

\paragraph{Probabilistic Approaches:}
The models by \cite{buyl_debayes_2020,li_toward_2022} are fitted using a prior that is informed of sensitive attributes to allow the rest of the model to focus on other aspects. 
When the model is used, the sensitive prior is replaced by one oblivious to the sensitive attributes. The intention is to produce fair recommendations along with neutral representations.

\cite{frisch_co-clustering_2021} explicitly model a variable for representing the contribution of the sensitive attributes instead of using a sensitive prior.
Ideally, this sensitive variable can soak up all sensitive information, leaving the rest of the model neutral. When recommending, the model drops the parts of the model that are dependent on the sensitive attribute.

\subsubsection{Utility-Based Fairness}
Utility-Based fairness optimization attempts to balance utility measures that involve ground truth consideration on a group or individual level. 
While only group-level optimizations are found among the in-processing methods, there is still significant variation in the considered approaches.

\cite{yao_beyond_2017} proposes four Utility-Based fairness metrics for recommender systems,
then adapt and applies each metric as loss terms in a matrix factorization approach. One of the metrics is similarly adapted by \cite{dickens_hyperfair_2020} as a probabilistic soft logic rule.

Numerous variations of straight-forward loss terms based on Group Utility-Based Fairness Interpretations are proposed for different models: the contextual bandit approach proposed by \cite{huang_fairness-aware_2021} penalizes differences in cumulative mean rewards of different sensitive groups. \cite{liu_dual_2022} and \cite{borges_f2vae_2022} both supplement adversarial approaches with Utility-Based fairness loss enhancement. \cite{liu_dual_2022} penalize the absolute differences in pairwise recommendation loss of different sensitive groups, while \cite{borges_f2vae_2022} penalize differences in reconstruction loss of a protected group and that of the average user in their variational autoencoder model. Finally, \cite{yao_personalized_2021} train personalized regularization weights based on the loss of a specific sensitive group to force the matrix factorization model to focus more on their achieved utility.

\cite{wan_addressing_2020} considers a unique recommender system setting where users and items are segmented into market segments based on sensitive groups and item groups and argue that the utility within the market segments should be similar. The proposed model applies loss terms that penalize error variation between user groups, item groups, and market segments. The authors also explore a market segment-level parity alternative by penalizing the variances of predicted ratings instead of errors.

\cite{li_leave_2021} propose a less direct way of encouraging the model to value the utility of non-mainstream user groups more by adding decoder components to their representations and corresponding loss terms for reconstructing the inputs. The intention is to provide the model with a stronger incentive for properly encoding all users and items, which in turn may mitigate issues with favouring the utility of mainstream user groups at the expense of everyone else. A similar goal is pursued by \cite{liu_self-supervised_2022}, who devise a set of auxiliary goals for encouraging their model to produce richer representations of all users. 

For the reciprocal setting, \cite{zheng_fairness_2018} proposes to consider both the utility of the user that receives the recommendation and the utility of the recommended users themselves. On a global level, this scheme balances the utility of two user groups based on the user's role in individual recommendations, i.e., reference users and users being recommended to reference users.

\subsubsection{Custom}
\cite{bobadilla_deepfair_2021} utilize empiric trends in the input data to design a set of indexes that represent users' and items' intrinsic \textit{sensitive value}. 
They further design a loss term to penalize recommending items to users if the index values differ significantly. 
Loss enhancement is also applied in the neighbourhood-based collaborative filtering model proposed by \cite{burke_balanced_2018} to balance the contribution of peers of different sensitive groups when recommending. Specifically, the added loss term penalizes the absolute difference of the model-specific user-to-user weights of different sensitive groups.

\subsection{Architecture and Method}
\subsubsection{Loss enhancement}
\paragraph{Matrix Factorization}
The early works of \citeauthor{kamishima_enhancement_2012,kamishima_efficiency_2013} are the earliest identified research that satisfies all acceptance criteria in this survey. The four publications \citep{kamishima_enhancement_2012,kamishima_efficiency_2013,kamishima_recommendation_2018,kamishima_considerations_2017} all propose matrix factorization models where the fairness aspects are modelled through different, but related loss terms. They all share the overall goal and fairness objective of ensuring statistical parity of recommendations. 
Additionally, all train different sets of parameters for the different sensitive groups they consider. 
In the first iteration,  \cite{kamishima_enhancement_2012}  propose a loss term that is an approximation of the mutual information of the rating and the sensitive attributes. Next, \cite{kamishima_efficiency_2013} introduces an efficiency improvement with alternative loss terms that penalize differing ratings per sensitive group averaged over all items or individually. 
The paper by \cite{kamishima_considerations_2017} considers similar loss terms, but through a implicit-feedback recommender system using ranking-based evaluation. 
It is noted that the approach has little effect on the ranking order of each user. 
Finally, \cite{kamishima_recommendation_2018} returns to rating-based recommender systems, introducing two new methods matching the first and second moment of the distributions for ensuring statistical parity. Both methods approximate rating distributions given the sensitive attribute with normal distributions, and then penalize Bhattacharyya distance \citep{bhattacharyya1943measure} and mutual information respectively.

Another early contribution was by \cite{yao_beyond_2017}, who argue for fairness definitions based on matching utility rather than Recommendation Parity. 
They propose four new Utility-Based \textit{unfairness} metrics that measure imbalances in how well the system recommends for different sensitive groups. Further, they devise loss terms based on these metrics and an additional parity-based metric to compare how well models trained with the different loss terms fare when evaluated using all metrics.

\cite{zheng_fairness_2018} is concerned with recommending matches after speed-dating, which is a reciprocal recommendation setting in the sense that consumers are recommended to other consumers. The model predicts user impression of speed-dates with different partners and considers a Custom utility metric based on the similarity of user's expectation of a partner and their impression of the speed-date partner. The utility metric is also used in the added loss term, which is designed to maximize the utility of both users in each potential match. The motivation is to balance the utility achieved by users being recommended for and the utility achieved by the users said users are recommended. Considering the utility of both involved users may also improve the overall success of this specific application, as mutual interest is ideal in a matchmaking scenario.

The approach proposed by \cite{wan_addressing_2020} is designed to address retail marketing bias by better balancing the achieved utility in different market segments. In particular, they define market segments based on sensitive user groups and attributes of models used in marketing different items, e.g., one segment may make out male users and items only marketed using female models. The proposed approach attempt to achieve similar utility within each distinct market segment by penalizing error variance between the different segments and other groupings. An alternative configuration is also considered where the model instead penalizes predicted rating variance, resulting in a Recommendation Parity Fairness Interpretation. 

The last identified loss enhancement-based matrix factorization approach was proposed by \cite{yao_personalized_2021}. The key idea of the model is to improve the utility of disadvantaged users through personalized regularization. This is achieved through a multi-step process that alternates between training for the general recommendation task while keeping the personalized regularization weights static and updating the same parameters based on the recommendation loss of the disadvantages.

\paragraph{Neighbourhood-based Collaborative Filtering}
\cite{burke_balanced_2018} propose enhancing the loss of a user-based collaborative filtering approach to encourage users' neighborhood of peers to be better balanced with respect the considered sensitive attributes. To this end, they devise a loss term that penalizes if the coefficients used for weighting influence of peers are skewed towards a specific group, i.e., the sum of male peer coefficients is greater than that of female peers. 

\paragraph{Neural Collaborative Filtering}
\cite{bobadilla_deepfair_2021} are unorthodox in terms of fairness definition and approach. 
They give each item a value based on the distribution of, e.g., the gender of users who like it, thus representing how \textit{gendered} the item is. 
The items are then used reversely to measure how gendered each \textit{user} is based on the items they like. 
The authors go on to penalize recommending items with a gendered value far from the user's.

\cite{li_leave_2021} aims to improve the utility of collaborative filtering models for users that are not mainstream. Their approach involves a factorization step that involves user and item representations, where the representations double up as the latent representations in two autoencoders. The autoencoders are added to encourage the model to properly encode all input information in the latent representations, and not neglect information only relevant to a subset of the users.

\paragraph{Bandit}
The only identified bandit approach was proposed by \cite{huang_fairness-aware_2021}, and is a contextual bandit method that penalizes differences in cumulative mean rewards of different sensitive groups. 
The authors construct a synthetic dataset 
for video recommendations and define a reward function that, for instance, rewards cases where the gender of the user and the video speaker is the same. 

\subsubsection{Probabilistic}
\paragraph{Graph}
\cite{buyl_debayes_2020} consider link prediction in social graphs and applies a probabilistic model for training graph representations based on the work by \cite{kang_conditional_2019}. 
They encode prior knowledge of the relations in the network using a prior term, which frees up the representations from encoding the same knowledge. 
\cite{buyl_debayes_2020} leverage this by designing priors that contain sensitive information to be used during training but replaced in the final recommendations. 
\cite{li_toward_2022} further adapt the approach for peer recommendation in online learning and introduce changes to the negative sampling used during training.

\paragraph{Probabilistic Model}
\cite{farnadi_fairness-aware_2018} and \cite{dickens_hyperfair_2020} both apply models based on Probabilistic Soft-Logic (PSL) \citep{bach_hinge_2017} for fairness-aware recommendation. PSL allows probabilistic models using human-interpretable first-order logic. 
Both models apply a base set of logical rules for the general recommendation task, e.g.,
\begin{align*}
 & \text{SIMILAR\_USER}(u_1,u_2)\land \text{RATING}(u_1,i)  \implies  \text{RATING}(u_2,i),\\
 & \text{SIMILAR\_ITEM}(i_1,i_2)\land \text{RATING}(u,i_1)      \implies \text{RATING}(u,i_2).
\end{align*}
\cite{farnadi_fairness-aware_2018} extend the model with fairness rules based on parity, e.g., one sensitive group's rating of an item implies the rating of a different sensitive group and vice versa. 
\cite{dickens_hyperfair_2020} consider courser parity-based rules and add others for encouraging equal utility for the different sensitive groups. 
Further, they allow modellers to adjust the importance of different fairness terms.
They also discuss using the model together with an arbitrary black-box model to inject fairness and interpretability, which can be thought of as a form of re-ranking. 

In the work by \cite{kamishima_model-based_2016}, two different graphical models are proposed for modelling ratings independent of a sensitive group membership. 
The models are realized as latent class models and optimized through the  Expectation-Maximization algorithm.

\cite{frisch_co-clustering_2021} propose using a latent block model for clustering users and items, then model a new rating-mean based on the associated cluster mean, individual variables for the item- and user-specific influences, and finally an item-specific variable that is controlled by the user's sensitive attribute. This final variable models the sensitive information, and is only used during training, similar to how informed priors are used in \cite{buyl_debayes_2020}. The model is optimized using variational inference.

\subsubsection{Algorithmic}
\paragraph{Neural Collaborative Filtering}
\cite{islam_mitigating_2019} explicitly subtract sensitive projections in user representations in a neural collaborative filtering model. 
They consider both scenarios where there is a single or multiple  binary sensitive attribute(s), e.g., male/female and young/senior. 
Some of the same authors \citep{islam_debiasing_2021} propose a more intricate approach where they utilize transfer learning to pre-train user representations and neural network weights in a non-sensitive recommendation setting. 
The user representations are then processed similarly as by \cite{islam_mitigating_2019} to be used in a sensitive recommendation setting. 
The non-sensitive settings considered are film recommendation and social media action recommendation, while the sensitive settings are occupation and college major recommendation, respectively. 
A parity-based loss term is applied in addition to the user representation processing to incorporate fairness in sensitive settings.

\cite{li_fairsr_2022} propose a fairness-aware sequential recommender system in which an integral part is to train item representations for representing the contextual information of the items and their relations. 
The authors use fairness-aware sampling when training said representations. Specifically, the sampling probability is set to adjust for any empirical skewness in how an item is preferred by different sensitive groups.

\paragraph{Graph}
The approach by \cite{rahman_fairwalk_2019} is designed to be used in reciprocal settings and tested on recommending peers in social networks while considering sensitive groups based on gender and race. 
The base representation algorithm performs random walks over the graph by sampling the next user among the current user's peers, i.e., the users the current user has a relation to in the observed data. 
Their fairness view is introduced by first sampling the peer's sensitive attribute uniformly, then sampling as usual from the qualified peers only.

\cite{xu_fair_2021} work with knowledge graph-based recommender systems. 
They propose training user representations of an auxiliary graph for representing sensitive attributes and their hidden relationships through a multi-layered neural network. 
This user representation is combined with that of the original recommender system in a linear transformation and then factorized with the item representation from the original recommender system. 
Additionally, an adversarial network is trained to classify sensitive attributes from the compound user representations and used to filter out said information. 
The purpose of the auxiliary graph representation is stated to be to improve the modelling of multiple sensitive attributes and their interactions.

\subsubsection{Adversarial}
\paragraph{Matrix Factorization}
\cite{resheff_privacy_2019} apply an adversarial gradient-based model to remove information like gender and age in the latent user factors. 
The authors list both privacy and fairness aspects as motivations for adversarial filtering.

\cite{li_towards_2021} adopt the approach proposed by \cite{bose_compositional_2019} using multiple different recommender system specific models, as opposed to the more general setting of link-prediction considered by \cite{bose_compositional_2019}. The approach is applied using four different models, covering matrix factorization and neural collaborative filtering. They further extend the approach by proposing a secondary option for training single filters for combinations of sensitive attributes, which is compared to the main approach of training one filter for each attribute and taking the mean of filtered representations to apply combinations.

\paragraph{Graph}
\cite{bose_compositional_2019} proposes to filter combinations of sensitive attributes from graph representations dynamically and considers both link-prediction and recommender system applications using different setups. 
They train one filter per sensitive attribute and combine filters by aggregating the representations processed by the filters. 
Each sensitive attribute is further assigned an adversarial for removing the traces of said sensitive attribute. 
A binary mask is sampled during training to simulate different users who want to restrict the use of different combinations of sensitive attributes. This mask is used in practice to activate the respective filters. 

\cite{wu_learning_2021} assume a graph-based perspective and that pre-trained user and item representations are provided. 
They suggest training filter functions for filtering out sensitive information from both representations and using these to build higher-order neighbourhood representations iteratively. 
For instance, the first-order neighbourhood representation of a user is based on the filtered representations of the items the user has liked or interacted with, 
the second-order neighbourhood contains the first-order neighbourhood representation of the same items, and so on. 
A multi-layered neural network is used to simultaneously process the first- and second-order neighbourhood representations into the final network level representation, 
with the motivation to capture higher-order sensitive information and how the different levels relate. Adversarial models are applied to both the filtered initial user representations and the final network-level user representations.

\cite{liu_self-supervised_2022,liu_mitigating_2022,liu_dual_2022} also apply neighbourhood representations, along with adversarial models for removing sensitive information in the base representations. 
However, they differ from \cite{wu_learning_2021}  as they consider end-to-end systems where the base representations are trained as part of the same model,
and  in using the highest-order representations explicitly as the final representations.
The three papers themselves differ in how they construct higher-order neighbourhood representations. 
\cite{liu_self-supervised_2022,liu_dual_2022} reduce the contribution of higher-order representations by dividing with a function that increases linearly in the order. 
\cite{liu_mitigating_2022} construct higher-order representations by passing the previous-order representations through a neural network and also explicitly considers the representations of neighbours that are increasingly further removed in the graph.
The approaches further differ in their application of additional fairness optimization. 
\cite{liu_dual_2022} propose two new loss terms: one for penalizing the covariance of the actual sensitive attribute and the one outputted by the adversarial model, and another for penalizing differences in pairwise losses of different sensitive groups. 
The former further enhances the neutrality of the representations, while the latter has an equality of utility motivation. 
.\cite{liu_self-supervised_2022} proposes to enhance the base representations by designing and applying a set of loss terms that encourage the representation of more complex information to mitigate the poor representation of underrepresented sensitive groups in the dataset.

\paragraph{Neural Collaborative Filtering}
A neural collaborative filtering model for fairness-aware news recommendation is proposed by \cite{wu_fairness-aware_2021}. 
The key idea is to contain all sensitive information in a part that can be disregarded when the model is applied, similar to the priors in \cite{buyl_debayes_2020}. 
Two separate user representations are trained: one is used for classifying the sensitive attribute and has the intention of aggregating sensitive information, while the other is designed for housing everything else and is coupled with an adversarial model to remove sensitive information. 
The sum of both user representations is used for recommendation during training while encouraging that the representations are orthogonal through adding a loss term. Only the neutral representation is used once the model is finished training.

\cite{closing_2022} propose to improve fairness in a job recommender system by filtering out gender information in representations of user resumes. To this end, they first train new word embeddings on resumes and job application texts within a proprietary dataset. The trained word embeddings are then used to encode the resume texts, and the encoded texts serve as inputs to their proposed neural recommender model. An adversarial model is applied to filter out gender information at a specific layer in the multi-layered neural network model. The authors also explore a simple alternative where they instead replace gendered words with neutral words before training the word embeddings. However, the adversarial approach is shown to outperform this alternative.

\cite{wu_selective_2022} follow \cite{bose_compositional_2019,li_towards_2021} in letting the users decide which sensitive attributes can be considered when generating recommendations. 
However, while the preceding research train multiple filters for different sensitive attributes that are dynamically plugged into the recommender system pipeline when activated, the filtering components in \cite{wu_selective_2022} are static parts of the model that dynamically change behaviour based on personalized prompts concatenated with the input. 
The filtering components are based on the \textit{adaptor} proposed by \cite{adaptor_2019} and trained along with different discriminators while keeping the remaining model parameters frozen.

The framework proposed by \cite{wei_comprehensive_2022} considers multiple Fairness Interpretations simultaneously. 
The framework consists of two main loops: an inner loop in which users are initialized with the latest parameters suggested by a meta-model and then optimized for different tasks, and an outer loop where the results of the inner loop are used to update the parameters of the meta-model to produce better user initializations in the next cycle. 
The framework applies two different adversarial models, which both attempt to detect sensitive attributes: the first one is fed user representations based on trained context representations and the users' observed ratings, while the second considers the predicted ratings, the corresponding observed ratings, and item representations. 

\paragraph{Variational Autoencoder}
\cite{borges_f2vae_2022} use a variational autoencoder (VAE) as their main model. 
The VAE is considered a collaborative recommender, where the decoded latent representation is interpreted as an encoding from which recommendations are extracted. 
The VAE is extended with an adversarial model for training the model to produce neutral latent representations and a loss term for encouraging the model to be equally good at reconstructing the inputs of a specific \textit{protected} sensitive group as at reconstructing the inputs of all users on average.

\section{Post-Processing Methods}
Post-processing methods share one of the main benefits of pre-processing methods in being flexible with respect to which recommender system model is used. Additionally, post-processing methods do not affect the raw data but are arguably the least flexible approaches when it comes to modelling since they are constrained by the provided recommendation and the data used to train the model. Post-processing methods are not as popular as in-processing methods but have received more attention than pre-processing methods.

\subsection{Fairness Optimization}
\begin{table}[htb]
\begin{center}
\caption{Overview of the identified post-processing approaches structured by the Fairness Interpretation and Fairness Incorporation of their optimization. Approaches that consider multiple Fairness Interpretations are listed in multiple rows.}\label{tab:fair_post}%
\begin{tabular}{p{0.27\textwidth}p{0.2\textwidth}p{0.4\textwidth}}
\toprule
& & \textbf{Re-ranking}\\
\midrule
\textbf{Recommendation Parity} & \textbf{Global} & \cite{dickens_hyperfair_2020}\newline\cite{ashokan_fairness_2021}\\
& & \\
& \textbf{Local} & \\
\midrule
\textbf{Neutral \newline Representation} & &\\
\midrule
\textbf{Utility-Based} & \textbf{Group} & \cite{ashokan_fairness_2021}\newline\cite{wu_multi-objective_2022}\\
& & \\
& \textbf{Individual} & \cite{wu_tfrom_2021}\\
\midrule
\textbf{Custom} & & \cite{edizel_fairecsys_2020}\newline\cite{paraschakis_matchmaking_2020}\newline\cite{patro_fairrec_2020}\newline\cite{patro_incremental_2020}\newline\cite{biswas_toward_2021}\newline\cite{lorenz_two_2021}\\
\bottomrule
\end{tabular}
\end{center}
\end{table}

\subsubsection{Recommendation Parity}
Both the post-processing techniques for Recommendation Parity included in the survey use the \textit{global} perspective. 
\cite{ashokan_fairness_2021} propose using results from the training data to align better the rating distribution of the two sensitive groups in a binary setting. To this end, they add the mean rating difference of the two sensitive groups during training to the predicted ratings of one of the groups when using the model for a new recommendation.

The approach in \cite{dickens_hyperfair_2020} discussed in Section \ref{sec:inproc:fairness:recparity} is also applicable as a re-ranker of a base recommender model. 
Thus, their proposed probabilistic soft logic rule comprises a second identified strategy for optimizing Global Recommendation Parity in post-processing methods.

\subsubsection{Utility-Based Fairness}

\paragraph{Group:}
The \textit{re-rating} approach proposed by \cite{ashokan_fairness_2021} considers a scheme where the per-item average rating error for each sensitive group, as observed in the training data, is added to the individual ratings of the users. This method is analogous to the previously covered parity scheme proposed in the same research.

The only identified approach that optimizes group Utility-Based fairness in a two-sided fairness approach is proposed by \cite{wu_multi-objective_2022}. Through applying the Frank-Wolfe algorithm \citep{frank_wolfe}, the authors optimize for consumer-side fairness by minimizing the variance of a differentiable definition of NDCG, along with the general recommendation and provider-side fairness objectives.

\paragraph{Individual:}
A multi-stakeholder approach is proposed by \cite{wu_tfrom_2021}, in which the consumer-side fairness objective is to fairly distribute the loss of utility incurred by the producer-side exposure considerations among the users. They devise a two-step approach, where the first step attempts to identify and fix highly preferred items that still have to reach their maximum exposure in the recommendation lists of the users. Each user is assigned one item at a time in a manner to even out the benefit of choosing first. The second step fills in the free recommendation slots with the items that still require exposure per the provider-side objective.

\subsubsection{Custom}
\cite{patro_fairrec_2020,patro_incremental_2020,biswas_toward_2021} all consider multi-stakeholder settings where the consumer-side objective is to distribute the loss of utility among users fairly. However, unlike \cite{wu_tfrom_2021}, they propose applying a utility definition that is not tied to the ground truth. 
Their shared utility measure is purely based on the preference values outputted by the original recommender system and produces values indicating how far from \textit{optimal} the new recommendations are deemed. \cite{patro_fairrec_2020}  and \cite{biswas_toward_2021} both propose similar two-step approaches as that of \cite{wu_tfrom_2021} but opt for guaranteeing the producer-side objective in the first step by allocating items in need of exposure in turn to users while attempting to prioritize items preferred by the user. The second step fills the remaining slots with the users' most preferred items. Finally, multi-sided fairness in a dynamic setting is explored by \cite{patro_incremental_2020} who attempts to retain both provider- and consumer-side fairness when facing different incremental changes to the recommendation, e.g., a gradual transition to a new base recommender model. Individual fairness is preserved by introducing lower-bound user utility constraints in the proposed integer linear programming model.

A similar setting is considered by \cite{lorenz_two_2021} whose approach also optimizes for two-sided fairness and applies custom user utility tied to the base recommender's outputted preference scores system. The approach positions itself to maximize the custom utility of worse-off users and items simultaneously in both regular and reciprocal recommendation settings.

\cite{edizel_fairecsys_2020}  focus on providing users recommendations that are uncorrelated with their sensitive attributes. 
The goal shares many parallels with in-processing approaches that optimize for intermediate representations that are uncorrelated with the sensitive attributes but operate on sets of recommendations due to the post-processing nature. 
The key idea is to allow users to inherit the recommendations of similar or arbitrary users belonging to different sensitive groups, thus muddying the correlation between the sensitive attributes and both recommendation lists and individual recommendations.

Another re-ranking approach is by \cite{paraschakis_matchmaking_2020}, which considers an individual fairness definition based on calibration by user preferences. 
Specifically, they consider a matchmaking setting where users specify how important it is for them to date within their race or religion. The problem is optimized through dynamic programming.

\subsection{Architecture and Method}
\subsubsection{Neighbourhood-Based Collaborative Filtering}
The approach proposed by \cite{ashokan_fairness_2021} is technically a re-\textit{rating} approach since they consider rating-prediction-based recommender systems but share many similarities with re-ranking approaches. The key idea is to attempt correct predictions of members of different sensitive groups by adding the average prediction errors for each item and sensitive group, as observed in the training set, to new predictions. 
A second parity-based option instead adds the average difference of the rating predictions given to different sensitive groups for each item.

\subsubsection{Matrix Factorization}
\cite{edizel_fairecsys_2020} focus on ensuring the generated top-$k$ recommendations are uncorrelated with sensitive groups by making users inherit the recommendations of other users belonging to a different sensitive group. 
While parts of the approach are enhanced when the same top-$k$ recommendations are recommended to multiple users by the base recommender system, the approach also works when all top-$k$ recommendations are unique, which is not unlikely for a large $k$ and a large item catalogue. 
Two different schemes for recommendation inheritance are evaluated: random and based on similarity. The former is shown to be more effective at reducing how indicative the recommended set is of the user's sensitive group but also reduces the utility quicker.

The works produced by \cite{patro_fairrec_2020,patro_incremental_2020,biswas_toward_2021} 
have an overlapping set of authors, and all consider the same two-sided recommendation setting. 
In \cite{patro_fairrec_2020}, the primary goal is to satisfy minimum requirements for provider-side exposure, subject to a secondary goal of distributing the loss of utility among the users fairly. 
They propose a two-step approach where first 
items are distributed among the user by allocating one item per user in turn to satisfy the items' minimum exposure criteria. 
The users are given the best remaining item according to said user's predicted preferences. 
Secondly, the remaining recommendation list of each user is filled based on the original recommendation. 
The approach is proven to guarantee recommendations that are \textit{envy-free up to one item} (EF1) \citep{envy_free}. 
The subsequent work by \cite{biswas_toward_2021} 
improves the model by identifying and removing \textit{envy-circles}, which are directed cycles in a graph representation of user envy, without affecting the EF1 guarantee.
\cite{patro_incremental_2020} look into incrementally updating the recommendation according to new data or major changes while retaining user and provider-side fairness. 
They consider three scenarios: switching the underlying recommender system, incorporating new data, and considering additional features in relevance estimation. 
For this approach, user-side fairness is the primary goal, and its performance is enforced through constraints in the proposed integer linear programming approach.

Similarly, \cite{lorenz_two_2021} also optimizes consumer- and provider-side fairness with respect to custom utility measures based on the base recommender's predictions and ranking exposure. Their key insight is to consider the utility of each user and item an objective, but ordering the objectives by performance for Pareto efficiency comparisons, e.g., the utility of users $u_1$ and $u_3$ are compared if they achieve the worst utilities in the two compared solutions. This objective formulation and ordering render Pareto efficiency equivalent to Lorenz efficiency, meaning that the utility of the worse-off users and items is maximized, and the model is optimized using the Frank-Wolfe algorithm. 

\cite{wu_tfrom_2021} considers a similar two-sided setting as the work above but opens for providers having more than one item each.
Further, the provider exposure is not considered uniform across recommendation ranks but higher in better positions on the recommended lists. For each recommendation position, the approach iterates through the users in an order based on the current recommendation utility and attempts to fill the position with a high-ranking item in the user's original recommendation, subject to maximum restrictions on provider exposure. Unfilled positions are since filled with items for meeting exposure requirements. 

Another two-sided approach is found in \cite{wu_multi-objective_2022}, which differs from the aforementioned two-sided approaches by relying more heavily on established optimization models. 
The consumer-side fairness objective is set to minimize the variance of smooth, i.e., differentiable, NDCG. In contrast, the provider-side fairness objective similarly considers the variance of exposure achieved by different items. 
The setup is applied by considering the base recommendation of multiple models, including two matrix factorization models and one neural collaborative filtering model, and produces multiple Pareto optimal solutions. 
The final solution is selected among these by favouring the solution that yields the best outcome for the worst-off objective.
Thus, unlike \cite{patro_fairrec_2020,wu_tfrom_2021} whose methods prioritized provide- and consumer-side fairness, respectively, through the order in which the objectives were considered, this approach does not explicitly favour one objective.

\subsubsection{Classification methods}
A return to matchmaking recommender systems is found in \cite{paraschakis_matchmaking_2020}, which details a calibration-based fairness approach. 
The approach considers the case where users define their preference for dating within their race or religion on a percentage scale, and the fairness objective is defined as making sure that the recommended list of each user has a racial or religious composition close to their preference. One proposed way to model the fair recommendation problem is to frame it as a Knapsack problem and find optimal solutions using dynamic programming. They also propose an alternative Tabu-search-based approach, which scales better but does not guarantee an optimal solution. The fairness evaluation is based on the same calibration applied throughout the approach as a fairness definition. 

\section{Metrics}\label{sec:metrics}
A plethora of different metrics have been proposed and applied in the identified research. Contributing factors to this great number of metrics are varying fairness definitions and the fact that different recommender system settings pose different requirements, e.g., rating vs ranking, binary vs multivalent sensitive attributes, and user-item vs user-user recommender systems. 
Another significant factor is how recently the topic has become relevant and the subsequent lack of consensus regarding evaluation. The metrics have all been structured according to the fairness categories they adhere to, and liberties have been taken in grouping similar metrics under new descriptive names. Further, formulas have been adapted and rewritten in the same notation for consistency. A lookup table for said notation can be found in Table \ref{tab:basic_notation}. Finally, each subsection covers a Fairness Interpretation and presents a table of the identified metrics, key contexts, and a list of the research that applied them.

\subsection{Recommendation Parity Metrics}
\begin{table}[htb]
\tiny
\begin{center}
\caption{Overview of identified Recommendation Parity metrics, their key properties and the research that has applied them.}\label{tab:parity_metrics}%
\begin{tabular}{p{0.15\textwidth}p{0.12\textwidth}p{0.1\textwidth}p{0.06\textwidth}p{0.06\textwidth}p{0.3\textwidth}}
\toprule
\textbf{Name} & \textbf{Fairness Subcategory}  & \textbf{Sensitive Attribute}  & \textbf{Rec. Dynamic} & \textbf{Rec. Type} & \textbf{Research}\\
\midrule
Item-level Parity & Local Parity & Binary\footnotemark[1] & User-Item & Mixed & \cite{bose_compositional_2019}\newline \cite{islam_debiasing_2021}\newline \cite{frisch_co-clustering_2021}\newline\cite{wu_learning_2021} \\
& & & & &\\
Item-level Rating Deviation & Local Parity & Multivalent   & User-Item  & Ranking & \cite{xu_fair_2021} \\
& & & & &\\
Global-level Parity  & Global Parity  & Binary  & User-Item  & Rating & \cite{yao_beyond_2017}\newline\cite{farnadi_fairness-aware_2018}\newline\cite{dickens_hyperfair_2020}\newline\cite{ashokan_fairness_2021}\newline\cite{fang_fairroad_2022}\newline\cite{closing_2022} \\
& & & & & \\
Mutual Information, Rating & Global Parity & Binary & User-Item & Rating & \cite{kamishima_enhancement_2012}\newline\cite{kamishima_efficiency_2013} \\
& & & & & \\
Kolmogorov-Smirnov & Global Parity & Binary & User-Item & Mixed & \cite{kamishima_model-based_2016}\newline\cite{kamishima_considerations_2017}\newline\cite{kamishima_recommendation_2018}\\
$\chi^2\text{-test}$ & Local Parity & Binary & User-item & Ranking & \cite{frisch_co-clustering_2021}\\
& & & & & \\
Group-to-group Variance & Global Parity & Multivalent & User-User & Ranking & \cite{rahman_fairwalk_2019}\newline\cite{buyl_debayes_2020}\\
& & & & & \\
Sensitive-group Share & Global Parity & Multivalent & User-User & Ranking & \cite{rahman_fairwalk_2019}\\
\bottomrule
\end{tabular}
\footnotetext[1]{Binary definition has also been applied to multivalent settings by applying it for all possible pairs.}
\end{center}
\end{table}

\subsubsection{Item-level Parity}\label{sec:item_level_parity}
Three identified metrics share a general design for summarizing the disparity of ratings or recommendations at the item level. All three metrics measure the item-level difference of ratings/recommendations aggregated by sensitive groups and a final aggregation by items. 
\begin{equation}\label{eq:comp_met}
    \hat{\mathbb{E}}_{v \in \mathcal{V}}\bigl[\arrowvert\hat{\mathbb{E}}_{u \in \mathcal{U}_{s_1}}[\hat{r}_{uv}]-\hat{\mathbb{E}}_{u \in \mathcal{U}_{s_2}}[\hat{r}_{uv}]\arrowvert\bigl]\nonumber
\end{equation}
\cite{bose_compositional_2019,wu_learning_2021} apply identical metrics that consider the simple absolute difference of item ratings for different groups in binary-sensitive attribute settings. \cite{bose_compositional_2019} also consider multivalent sensitive attributes, for which Equation \ref{eq:comp_met} is expanded to consider all possible pairs of sensitive groups.

\begin{equation}\label{eq:debias_met}
    \hat{\mathbb{E}}_{v \in \mathcal{V}}\left[\left\arrowvert \ln{\left(\hat{\mathbb{E}}_{u \in \mathcal{U}_{s_1}}\left[1\{\hat{y}_{uv}\}\right]\right)} - \ln{\left(\hat{\mathbb{E}}_{u \in \mathcal{U}_{s_2}}\left[1\{\hat{y}_{uv}\}\right]\right)}\right\arrowvert\right]\\
\end{equation}

\begin{equation}\label{eq:coclust_met}
    \max_{v1,v2 \in \mathcal{V} \arrowvert v1\neq v2}\left\arrowvert \left(\hat{\mathbb{E}}_{u \in \mathcal{U}_{s_1}}\left[1\{\hat{r}_{uv_1} > \hat{r}_{uv_2}\}\right]\right) - \left(\hat{\mathbb{E}}_{u \in \mathcal{U}_{s_2}}\left[1\{\hat{r}_{uv_1} > \hat{r}_{uv_2}\}\right]\right)\right\arrowvert\\
\end{equation}

\cite{islam_debiasing_2021,frisch_co-clustering_2021} both define similar concepts named $\epsilon$-(differentially)fair, where individual $\epsilon$'s reflect how much the recommendation of a single item differs in a binary sensitive group setting. The former considers the probability of recommending items to different sensitive groups, Equation \ref{eq:debias_met}, while the latter considers the probability of ranking an item higher than another item to different sensitive groups, Equation \ref{eq:coclust_met}. \cite{islam_debiasing_2021} takes inspiration from \textit{differential privacy} \citep{Dwork2011}, and subsequently have logarithmic terms in the absolute difference, but opt for computing the average $\epsilon$ and not the maximum. \cite{frisch_co-clustering_2021} does not cite differential metrics or concepts, but is concerned with the maximum $\epsilon$ and not the average. Out of the two aggregations, the maximum poses a stronger guarantee than the average and is more in line with the differential definition.

\subsubsection{Item-level Rating Deviation}
\begin{align}\label{eq:fair_met}
    \hat{\mathbb{E}}_{v \in \mathcal{V}}\sqrt{\hat{\mathbb{E}}_{j \in \mathcal{S}}\left[\left(\hat{\mathbb{E}}_{u \in \mathcal{U}_j}[\hat{r}_{uv}]-\mu_v\right)^2\right]},\nonumber\\
    \text{where } \mu_v = \hat{\mathbb{E}}_{k \in \mathcal{S}}\bigl[\hat{\mathbb{E}}_{u \in U_k}[\hat{r}_{uv}]\bigl].\nonumber
\end{align}
While the Item-level Parity metrics consider the mean difference of predicted item ratings between sensitive groups, \cite{xu_fair_2021} opt for measuring the mean standard deviation of the same ratings. The metric is inherently capable of considering more than two sensitive groups, and the square difference term leads to larger penalties for large differences than the corresponding penalties in the metrics applying the absolute difference instead.

\subsubsection{Global-level Parity}
\begin{equation}
    \left\arrowvert\hat{\mathbb{E}}_{v \in \mathcal{V}}\left[\hat{\mathbb{E}}_{u \in \mathcal{U}_{s_1}}\left[\hat{r}_{uv}\right]\right] - \hat{\mathbb{E}}_{v \in \mathcal{V}}\left[\hat{\mathbb{E}}_{u \in \mathcal{U}_{s_2}}\left[\hat{r}_{uv}\right]\right]\right\arrowvert\nonumber
\end{equation}
The Global-level Parity metric is applied in multiple research and is simply the absolute difference of the mean predicted rating or preference score of different sensitive groups.

\subsubsection{Mutual Information, Rating}\label{sec:mutual_information}
\begin{equation}
    \sum_{s \in \mathcal{S}}\int \text{P}(\hat{r}, s)\log \frac{\text{P}(\hat{r}\arrowvert s)}{\text{P}(\hat{r})} d\hat{r}\nonumber
\end{equation}
Mutual information is a concept from information theory, comprising a measure of the mutual dependency of two variables. It has been applied in fair recommender systems to measure the mutual dependency of the rating and the sensitive attribute. Due to the probabilistic definition, non-probabilistic models applying the metric have resorted to different methods for approximating the measure, e.g., empiric approximations of probabilities and bucketing the ratings into intervals to replace the inner integral with a sum.

\subsubsection{Kolmogorov-Smirnov Statistic}
\begin{equation}
    \sup_{\hat{r}} \arrowvert F_{s_1}(\hat{r}) - F_{s_2}(\hat{r}) \arrowvert\nonumber
\end{equation}
The Kolmogorov-Smirnov(KS) statistic measures how different two probability distributions are, and its estimation involves the cumulative distributions of the probability distributions. Both cumulative distributions are typically empirically defined based on the outputted ratings for different sensitive groups when used in recommender systems. $\sup$ is the supremum, meaning that the statistic returns the upper-bounded difference between the distributions of the ratings.

\subsubsection{\texorpdfstring{$\chi^2$}  --Test}
$\chi^2$-tests are typically used to determine if the difference in collections of categorical data is probable, given that they were sampled from the same distribution. \cite{frisch_co-clustering_2021} applied $\chi^2$-test to test the independence of group membership and user gender. Since group membership influences the rating within their Latent Block Model, the groups' gender composition should ideally reflect the overall gender composition when pursuing recommendation parity.

\subsubsection{Group-to-group Variance}
\begin{align}
    \hat{\mathbb{E}}_{a,b \in \mathcal{S}\times\mathcal{S}\arrowvert a \neq b}[(&N_a - N_b)^2],\nonumber\\
    \text{where }&N_a = N_{s_i, s_j} =  \frac{\arrowvert \{\hat{y}_{u_1,u_2}\arrowvert u_1\in\mathcal{U}_{s_i},u_2\in\mathcal{U}_{s_j}\}\arrowvert}{\arrowvert \mathcal{U}_{s_i}\times\mathcal{U}_{s_j}\arrowvert}. \nonumber
\end{align}
In reciprocal recommendations, each recommendation involves two users who may belong to different sensitive groups. The Group-to-group Variance metric considers the variance of the acceptance rate, i.e., recommendation rate, of different combinations of sensitive groups.

\subsubsection{Sensitive-group Share}
\begin{equation}
    \mathcal{S}\text{-share}(s) = \frac{1}{\arrowvert \mathcal{S} \arrowvert}-\frac{\sum_{u_1 \in \mathcal{U}}\sum_{u_2\in \mathcal{U}_s}\frac{1\{u_2 \in \text{Rec}_{u_1}\}}{\arrowvert \text{Rec}_{u_1}\arrowvert}}{\arrowvert \mathcal{U}\arrowvert}\nonumber
\end{equation}
The Sensitive-group Share metric measures how well an individual sensitive group $s$ is represented in the recommendations of all users in a reciprocal setting. It subtracts the real representation ratio for said group from the ideal uniform ratio, such that the output represents how far from ideal the recommendations are with respect to single sensitive groups.

\subsection{Neutral Representation Metrics}
Neutral Representation fairness is a special case in that it does not explicitly concern itself with the actual outputs of the model. This also extends to the metrics of the Fairness Interpretation. However, while some research adapting this Fairness Interpretation in their optimization only evaluate how neutral the representations are, others adapt metrics from other interpretations to evaluate the recommendation explicitly.
\begin{table}[htb]
\tiny
\begin{center}
\caption{Overview of identified Neutral Representation metrics, their key properties and the research that has applied them.}\label{tab:rep_metrics}%
\begin{tabular}{p{0.2\textwidth}p{0.15\textwidth}p{0.06\textwidth}p{0.06\textwidth}p{0.35\textwidth}}
\toprule
\textbf{Name} & \textbf{Sensitive groups} & \textbf{Rec. Dynamic} & \textbf{Rec. Type} & \textbf{Research}\\
\midrule
Sensitive Reclassification & Multivalent & Mixed & Mixed & \cite{bose_compositional_2019}\newline\cite{resheff_privacy_2019}\newline\cite{buyl_debayes_2020}\newline\cite{li_towards_2021}\newline\cite{wu_learning_2021}\newline\cite{li_toward_2022}\newline\cite{borges_f2vae_2022}\newline\cite{wu_selective_2022}\newline\cite{wei_comprehensive_2022}\\
\bottomrule
\end{tabular}
\end{center}
\end{table}

\subsubsection{Sensitive Reclassification}
The vast majority of research optimizing for sensitive neutral representations perform some form of evaluation of how well sensitive information can be reclassified through the representations. This evaluation is usually performed by training an auxiliary classification model specifically for identifying the sensitive attributes of users given their representations, and the re-classification score becomes an inverse measure of how well sensitive information has been eliminated. \textit{Accuracy}, \textit{F1 score} and \textit{Area Under the ROC Curve}(AUC) are all metrics that have been used for this purpose. AUC is the total area under the curve of the curve you get by plotting the true positive rate and the false positive rate while moving the threshold used to split positive and negative classifications. AUC is by far the most applied classification metric.

\subsection{Utility-Based Fairness}
Utility-Based fairness metrics are specifically tied to utility functions that consider the \textit{ground truth} in their definitions, i.e., the utility or utility contribution of a single user increases if an item in the test set is successfully recommended for said user. This covers established pure-recommendation metrics, as well as other specialized utility measures.
\begin{table}[htb]
\tiny
\begin{center}
\caption{Overview of identified Utility-Based Fairness metrics, their key properties and the research that has applied them.}\label{tab:util_metrics}%
\begin{tabular}{p{0.15\textwidth}p{0.12\textwidth}p{0.1\textwidth}p{0.06\textwidth}p{0.06\textwidth}p{0.3\textwidth}}
\toprule
\textbf{Name} & \textbf{Fairness Subcategory} & \textbf{Sensitive groups} & \textbf{Rec. Dynamic}& \textbf{Rec. Type} & \textbf{Research}\\
\midrule
Group Rating Error Difference & Group & Binary\footnotemark[1] & User-item & Mixed & \cite{yao_beyond_2017}\newline\cite{farnadi_fairness-aware_2018}\newline\cite{dickens_hyperfair_2020}\newline\cite{islam_debiasing_2021}\newline\cite{wu_learning_2021}\newline\cite{fang_fairroad_2022}\\
& & & & & \\
Group Rating Error Deviation & Group & Multivalent & User-item & Mixed & \cite{xu_fair_2021}\\
& & & & & \\
Group Utility Difference & Group & Binary\footnotemark[1] & Mixed & Mixed & \cite{zheng_fairness_2018}\newline\cite{islam_mitigating_2019}\newline\cite{huang_fairness-aware_2021}\newline\cite{liu_dual_2022}\newline\cite{borges_f2vae_2022}\newline\cite{liu_mitigating_2022}\newline\cite{liu_self-supervised_2022}\newline\cite{wei_comprehensive_2022}\\
& & & & & \\
Utility Variance & Mixed & Multivalent & User-item & Mixed & \cite{rastegarpanah_fighting_2019}\newline\cite{wu_tfrom_2021}\newline\cite{wu_multi-objective_2022}\\
Utility Delta & Group & Multivalent & User-item & Mixed & \cite{li_leave_2021}\newline\cite{slokom_towards_2021}\\
& & & & & \\
Mutual Information, Relevance & Group & Binary & User-Item & Ranking & \cite{kamishima_considerations_2017}\\
& & & & & \\
Inequality of Odds & Group & Binary & User-item & Ranking & \cite{li_toward_2022} \\
& & & & & \\
Inequality of Opportunity & Group & Binary & User-item & Ranking & \cite{kamishima_considerations_2017}\\
& & & & & \\
Protected Utility & Group & Binary & User-item & Rating & \cite{yao_personalized_2021}\\
& & & & & \\
GEI & Group & Binary & User-item & Rating & \cite{ashokan_fairness_2021}\\
& & & & & \\
Generalized Cross Entropy & Group & Multivalent & User-item & Ranking & \cite{liu_dual_2022}\\
& & & & & \\
F-Statistic Utility & Group & Multivalent & User-item & Rating & \cite{wan_addressing_2020}\\
\bottomrule
\end{tabular}
\footnotetext[1]{Binary definition, but has also been applied to multivalent settings by taking the mean or standard deviation of all possible pairs/one-vs-all values.}
\end{center}
\end{table}

\subsubsection{Group Rating Error Difference}\label{sec:beyond}
\begin{align}
    &\text{ValueUnfairness}=\nonumber\\& \phantom{===}
    \hat{\mathbb{E}}_{v \in \mathcal{V}} \left[\biggl\arrowvert 
    \hat{\mathbb{E}}_{u \in \mathcal{U}_{s_1}}[\hat{r}_{uv} - r_{uv}] - 
    \hat{\mathbb{E}}_{u \in \mathcal{U}_{s_2}}[\hat{r}_{uv} - r_{uv}]\biggl\arrowvert\right]\label{eq:val_unfair}\\
    &\text{AbsoluteUnfairness}=\nonumber\\& \phantom{===}
    \hat{\mathbb{E}}_{v \in \mathcal{V}} \left[\biggl\arrowvert 
    \Bigl\arrowvert\hat{\mathbb{E}}_{u \in \mathcal{U}_{s_1}}[\hat{r}_{uv} - r_{uv}]\Bigl\arrowvert - 
    \Bigl\arrowvert\hat{\mathbb{E}}_{u \in \mathcal{U}_{s_2}}[\hat{r}_{uv} - r_{uv}]\Bigl\arrowvert\biggl\arrowvert\right]\label{eq:abs_unfair}\\
    &\text{OverestimationUnfairness}=\nonumber\\& \phantom{===}
    \hat{\mathbb{E}}_{v \in \mathcal{V}} \left[\biggl\arrowvert 
    \max\Bigl(0,\hat{\mathbb{E}}_{u \in \mathcal{U}_{s_1}}[\hat{r}_{uv}-r_{uv}]\Bigl) - 
    \max\Bigl(0,\hat{\mathbb{E}}_{u \in \mathcal{U}_{s_2}}[\hat{r}_{uv}-r_{uv}]\Bigl)\biggl\arrowvert\right]\nonumber\\
    &\text{UnderestimationUnfairness}=\nonumber\\ &\phantom{===}
    \hat{\mathbb{E}}_{v \in \mathcal{V}} \left[\biggl\arrowvert \max\Bigl(0,\hat{\mathbb{E}}_{u \in \mathcal{U}_{s_1}}[r_{uv}-\hat{r}_{uv}]\Bigl) - 
    \max\Bigl(0,\hat{\mathbb{E}}_{u \in \mathcal{U}_{s_2}}[r_{uv}-\hat{r}_{uv}]\Bigl)\biggl\arrowvert\right]\nonumber
\end{align}

These four metrics were all defined by \cite{yao_beyond_2017} and make out some of the few metrics that have gained a semblance of recognition in subsequent work on consumer-side recommendation fairness. Every metric accumulates absolute differences in per-item errors of two sensitive groups, and all of them involve mechanics for cancelling errors in case the model is similarly underperforming for both groups. The latter two focus on over- and under-estimation, respectively, while the former two consider both error types concurrently while differing in how the errors can cancel out. Specifically, Value Unfairness, Equation \ref{eq:val_unfair}, allows errors of the same type to cancel out, while Absolute Unfairness, Equation \ref{eq:abs_unfair}, allows all errors to cancel out regardless of error type.

\begin{equation}
    \hat{\mathbb{E}}_{v \in \mathcal{V}} \left[\biggl\arrowvert \hat{\mathbb{E}}_{u \in \mathcal{U}_{s_1}}\Bigl[\arrowvert\hat{r}_{uv} - r_{uv}\arrowvert\Bigl] - \hat{\mathbb{E}}_{u \in \mathcal{U}_{s_2}}\Bigl[\arrowvert\hat{r}_{uv} - r_{uv}\arrowvert\Bigl]\biggl\arrowvert\right]\label{eq:wu_learning}\\
\end{equation}
The metric applied in \cite{wu_learning_2021} resembles Absolute Unfairness, but considers the absolute errors at the user level instead of at the group level. One implication of this is a higher incurred penalty when the errors of the predicted ratings vary a lot within one or both groups, as they are not evened out by taking the group-wise mean before the absolute error is calculated, which is the case for Absolute Unfairness.

\subsubsection{Group Rating Error Deviation}
\begin{align}\label{eq:fair_met2}
    \hat{\mathbb{E}}_{v \in \mathcal{V}}\left[\sqrt{\hat{\mathbb{E}}_{i \in \mathcal{S}}\left[\left(\hat{\mathbb{E}}_{u \in \mathcal{U}_{i}}\left[\arrowvert\hat{r}_{uv} - r_{uv}\arrowvert\right]-\mu_v\right)^2\right]}\right],\nonumber\\
    \text{where } \mu_v = \hat{\mathbb{E}}_{j \in S}\left[\left(\hat{\mathbb{E}}_{u \in U_{j}}\left[\arrowvert\hat{r}_{uv} - r_{uv}\arrowvert\right]\right)\right].  \nonumber
\end{align}
Along with Item-level Rating Deviation, \cite{xu_fair_2021} propose another metric that considers the standard deviation of item-level statistics of different sensitive groups. This time, it is a Utility-Based fairness metric, with a central term considering the squared difference of mean user-level absolute item rating errors. While the metric is structurally similar to Group Rating Error, see in particular Equation \ref{eq:wu_learning}, it is important to note that this metric measures the mean \textit{standard deviation} of the group-wise rating errors instead of the mean \textit{difference} of the same errors. Further, the Group Rating Error Deviation metric is inherently capable of considering multivalent sensitive attributes.

\subsubsection{Group Utility Difference}
\begin{equation}
    \arrowvert \text{Util}(\mathcal{U}_{s_1}) - \text{Util}(\mathcal{U}_{s_2})\arrowvert\nonumber
\end{equation}
Among the considered research, some explicitly calculate the absolute difference of Recall, Precision and NDCG \citep{ndcg} achieved by different sensitive groups. Also included in this group of metrics are more implicit comparisons of utility metrics through tables or graphs, which has been observed considering MAE, cumulative mean custom rewards and custom matchmaking utility. A special case of this metric is applied in the reciprocal setting of \cite{zheng_fairness_2018}, 
where the users receiving the recommendation comprise one sensitive group, and the users that make up the recommended entities comprise the other sensitive group.

\subsubsection{Utility Variance}
\begin{align}
    &\hat{\mathbb{E}}_{u_1\in\mathcal{U}}\left[\hat{\mathbb{E}}_{u_2\in\mathcal{U}\arrowvert u_2\neq u_1}\left[(\text{Util}(u_1) - \text{Util}(u_2))^2\right]\right]\nonumber\\
    &\hat{\mathbb{E}}_{s_1\in\mathcal{S}}\left[\hat{\mathbb{E}}_{s_2\in\mathcal{S}\arrowvert s_2\neq s_1}\left[(\text{Util}(\mathcal{U}_{s_1}) - \text{Util}(\mathcal{U}_{s_2}))^2\right]\right]\nonumber
\end{align}
The variation of the utility of individual users or sensitive groups has been applied as fairness metrics. It is particularly suited for representing the utility spread when there are too many scores to cover individually, i.e., many users or sensitive groups. The identified variations of Utility Variance have centred around the utility metrics Mean Squared Error and NDCG.

\subsubsection{Utility Delta}
\begin{align}
    \Delta _s &=\text{Util}(\mathcal{U}_{s}) - \text{Util}_\text{org}(\mathcal{U}_{s})\nonumber\\
    \Delta _\text{diff} &= \arrowvert \Delta _{s_1} - \Delta _{s_2} \arrowvert\label{eq:delta_diff}
\end{align}
The Utility Delta metrics consider how the utility of specific sensitive groups changes when a fairness-aware model is compared to baselines. From a Utility-Based fairness view, a decrease in utility for a dominant sensitive group may be worth a subsequent increase in the utility of worse-off sensitive groups. Having measures of how the utility of specific user groups has changed is useful when considering such trade-offs. \cite{slokom_towards_2021} also consider the absolute difference of the $\Delta$-s of two sensitive groups to capture the dissymmetry in the magnitude of the changes, Equation \ref{eq:delta_diff}.

\subsubsection{Mutual Information, Relevance}
\cite{kamishima_considerations_2017} applies mutual information in a Utility-Based fairness evaluation of their ranking-based recommender system. The definition is structurally identical to that of the Recommendation Parity metric in Section \ref{sec:mutual_information}. The main difference is that the predicted rating variable is replaced with a binary relevancy variable, i.e., the degree of independence between relevant recommendations and the sensitive attribute is measured. 

\subsubsection{Inequality of Odds}
\begin{equation}
    \max(\arrowvert \text{FPR}_{s_1} - \text{FPR}_{s_2}\arrowvert, \arrowvert \text{TPR}_{s_1} - \text{TPR}_{s_2}\arrowvert)\nonumber
\end{equation}
Equality of Odds requires the true positive and false positive rates of different sensitive groups to be equal. \cite{li_toward_2022} propose a metric based on this definition for measuring the extent of potential infractions. The same research also applies Absolute Between-ROC Area(ABROCA), proposed in \cite{gardner_evaluating_2019}, which inherently considers all possible positive-boundary thresholds as opposed to a single fixed threshold.

\subsubsection{Inequality of Opportunity}
\begin{equation}
    \text{TPR}_{s_1} - \text{TPR}_{s_2}\nonumber
\end{equation}
\cite{kamishima_considerations_2017} considers a metric based on Equality of Opportunity, which requires equality of true positive rates of different sensitive groups. They opt for measuring infractions with a regular difference to avoid abstracting away the orientation of potential imbalances. 

\subsubsection{Protected Utility}
The protected utility metrics only consider the utility of specific sensitive groups that are considered protected. \cite{yao_personalized_2021} is the only identified research that adopts this metric type, and they use RMSE as their utility measure.

\subsubsection{Generalized Entropy Index}
\begin{equation}
    \text{GEI}(\alpha) = 
    \begin{cases}
    \frac{1}{\arrowvert \mathcal{S}\arrowvert \alpha(1-\alpha)}\sum_{s \in \mathcal{S}}\left(\left(\frac{\text{Util}(\mathcal{U}_s)}{\text{Util}(\mathcal{U})}\right)^\alpha - 1\right),& \alpha \neq 0,1\\
    \frac{1}{\arrowvert \mathcal{S}\arrowvert}\sum_{s \in \mathcal{S}} \frac{\text{Util}(\mathcal{U}_s)}{\text{Util}(\mathcal{U})}\ln\frac{\text{Util}(\mathcal{U}_s)}{\text{Util}(\mathcal{U})},& \alpha = 1\\
    -\frac{1}{\arrowvert \mathcal{S}\arrowvert}\sum_{s \in \mathcal{S}} \ln\frac{\text{Util}(\mathcal{U}_s)}{\text{Util}(\mathcal{U})},& \alpha = 0\nonumber\\
    \end{cases}
\end{equation}
Generalized Entropy Index(GEI) is a metric often used for measuring income inequality and has been applied both for measuring individual inequality and inequality among larger user groups in recommender systems. In the case of GEI based on individuals, $s$ is interpreted as each user having their own sensitive group $\mathcal{U}_s$. Note that the applied utility measure typically will average over all represented users to get comparative scores when calculating the utility of all users and the one based on a subset of the users. \cite{farnadi_fairness-aware_2018} consider GEI for both $\alpha=1$ and $\alpha=2$, where the former yields a special case known as the Theil index.

\subsubsection{Generalized Cross Entropy}
\begin{equation}
    \text{GCE}(m, S) = \frac{1}{\beta\cdot(1-\beta)}\left(\sum_{s \in S} \text{P}_f^{\beta}(s)\cdot \text{P}_m^{(1-\beta)}(s) -1\right)\nonumber
\end{equation}
Generalized Cross Entropy allows the evaluator to adjust the importance given to the utility of different sensitive groups, e.g., in settings where you have premium users or known marginalized users for whom one wants to boost the performance. $\text{P}_f$ is a fair probability distribution, e.g., a uniform distribution if the motive is to give equal focus to all sensitive groups. $\text{P}_m$ is a probability distribution over the model $m$'s utility. 

\subsubsection{\texorpdfstring{$F$}--Statistic}
The $F$-Statistic is typically calculated as part of a $F$-test used in analyses of variance. \cite{wan_addressing_2020} applies the $F$-statistic while considering the variance of rating errors between market segments and within them as a measure of fairness, where lower $F$-statistics indicate higher levels of fairness.

\subsection{Custom Fairness}
Custom fairness encompasses all definitions and measures that do not strictly consider Recommendation Parity, representation neutrality or Utility-Based fairness. Unlike Utility-Based fairness metrics, any utility considered in these metrics is not affected by the \textit{ground truth}. Additionally, while some of these metrics consider parity-based concepts, it concerns aspects other than recommendation, i.e., not explicitly rating, preference or ranking. For instance, parity with respect to derived/contextual user attributes or a particular item group.
\begin{table}[htb]
\tiny
\begin{center}
\caption{Overview of identified Custom fairness metrics, their key properties and the research that has applied them.}\label{tab:custom_metrics}%
\begin{tabular}{p{0.2\textwidth}p{0.15\textwidth}p{0.06\textwidth}p{0.06\textwidth}p{0.35\textwidth}}
\toprule
 & \textbf{Sensitive groups} & \textbf{Rec. Dynamic} & \textbf{Rec. Type} & \textbf{Research}\\
\midrule
Normalized Ranking Change & Multivalent & User-item & Ranking & \cite{patro_fairrec_2020}\newline\cite{patro_incremental_2020}\newline\cite{biswas_toward_2021}\\
& & & \\
Ranking Change Envy & Multivalent & User-item & Ranking & \cite{patro_fairrec_2020}\newline\cite{biswas_toward_2021}\\
& & & \\
Gini Coefficient, Preference & Multivalent & Mixed & Ranking & \cite{lorenz_two_2021}\\
& & & \\
Protected Item-group Recommendation Parity & Binary & User-item & Ranking & \cite{burke_balanced_2018}\\
& & & \\
Preferential Calibration & Multivalent & User-item & Ranking & \cite{paraschakis_matchmaking_2020}\\
& & & \\
Intrinsic Sensitive Attribute Match & Binary & User-item & Ranking & \cite{bobadilla_deepfair_2021}\\
& & & \\
Sensitive Neutral Recommended Items & Multivalent & User-item & Ranking & \cite{li_fairsr_2022}\\
& & & \\
Segment Recommendation Frequency Parity & Multivalent & User-item & Ranking & \cite{wan_addressing_2020}\\
& & & \\
Sensitive Reclassification, Pre-/Post- & Multivalent & User-item & Mixed & \cite{edizel_fairecsys_2020}\newline\cite{wu_fairness-aware_2021}\newline\cite{slokom_towards_2021}\\
\bottomrule
\end{tabular}
\end{center}
\end{table}

\subsubsection{Normalized Ranking Change}
\begin{align}
    \text{PrefUtil}_{u}&(\text{Rec}) = \frac{\sum_{v \in \text{Rec}} \text{pref}_{uv}}{\sum_{v' \in \text{Rec}_\text{org\_u}} \text{pref}_{uv'}}\label{eq:pref_util}\\
    \text{NormRankChg}&\text{Mean} = \hat{\mathbb{E}}_{u \in \mathcal{U}}\left[\text{PrefUtil}_{u}(\text{Rec}_u)\right]\nonumber\\
    \text{NormRankC}&\text{hgStd} = \nonumber\\
    &\sqrt{\hat{\mathbb{E}}_{u_1\in\mathcal{U}}\left[\hat{\mathbb{E}}_{u_2\neq u_1}\left[\left(\text{PrefUtil}_{u_1}(\text{Rec}_{u_1}) - \text{PrefUtil}_{u_2}(\text{Rec}_{u_2})\right)^2\right]\right]}\nonumber
\end{align}
\cite{patro_fairrec_2020,patro_incremental_2020,biswas_toward_2021} all consider a two-sided fairness setting and interpret the consumer-side fairness as how far the recommendations deviate from the original ranking when producer-side fairness has been taken into account. I.e., they use the intermediate preference scores of the re-ranked top-$k$ recommendations normalized by the optimal preference scores of the original top-$k$ recommendations as a proxy of the utility. This utility's mean and standard deviation are considered in their fairness evaluation.

\subsubsection{Ranking Change Envy}
\begin{align}
    \text{RankingEnvy}(u, u') &= \max(\text{PrefUtil}_{u}(\text{Rec}_{u'}) - \text{PrefUtil}_{u}(\text{Rec}_{u}), 0)\nonumber\\
    \text{RankChgEnvy} &= \hat{\mathbb{E}}_{u_1 \in \mathcal{U}}\left[\hat{\mathbb{E}}_{u_2 \neq u_1} \left[\text{RankingEnvy}(u_1, u_2)\right]\right]\nonumber
\end{align}
In multi-sided fairness settings, it is not uncommon to have a situation where the utility of one user may be increased by giving said users the recommendations given to another user instead of their own. The term \textit{Envy} has been taken to refer to cases where users would fare better given the recommendations of other users. Envy can arise in multi-sided recommendendation fairness when auxiliary considerations affect some users more than others, e.g., considerations of provider-side fairness. \cite{patro_fairrec_2020} and \cite{biswas_toward_2021} consider the aggregated envy of their proposed \textit{preference utility}, Equation \ref{eq:pref_util}.

\subsubsection{Gini Coefficient, Preference}
\begin{align}
    \text{PrefUtilExp}(u) &= \sum_{v \in \mathcal{V}}\text{pref}_{uv}\boldsymbol{P}_{uv} \boldsymbol{w}\label{eq:pref_util_exp}\\
    \text{GiniPref} &= \frac{\sum_{u_1,u_2 \in \mathcal{U} \times \mathcal{U}}\arrowvert \text{PrefUtilExp}(u_1) - \text{PrefUtilExp}(u_2)\arrowvert}{2 \arrowvert\mathcal{U}\arrowvert^2 \mathbb{E}_{u \in \mathcal{U}}[\text{PrefUtilExp}(u)] }\nonumber
\end{align}
\cite{lorenz_two_2021} propose adapting the Gini coefficient, frequently used to measure wealth inequality, to measure individual consumer fairness in their two-sided fairness setting. The Gini coefficient is measured for a custom utility measure based on the outputted preference scores of their base recommender system and the ranking positions of the re-ranked recommendations. Their proposed utility measure is defined in Equation \ref{eq:pref_util_exp}, where $\boldsymbol{P_{uv}}$ is a row vector of probabilities for recommending user $u$ item $v$ in different ranking positions, and $\boldsymbol{w}$ is a column vector of exposure/rank weights for the same ranking positions. 

\subsubsection{Protected Item-group Recommendation Parity}
\begin{equation}
   \frac{\sum_{u \in \mathcal{U}_{s_1}}\sum_{v \in \text{Rec}_{u}}\frac{\gamma (v)}{\arrowvert \mathcal{U}_{s_1}\arrowvert}}{\sum_{u' \in \mathcal{U}_{s_2}}\sum_{v' \in \text{Rec}_{u'}}\frac{\gamma (v')}{\arrowvert \mathcal{U}_{s_2}\arrowvert}} \nonumber
\end{equation}
Here $\gamma$ is a function that returns ``1" if the item belongs to any protected item groups and ``0" if none of its item groups is protected, and the metric measures how balanced the recommendation of the protected item groups are between two sensitive groups, i.e., a value of 1 is optimal. \cite{burke_balanced_2018} applied this metric to evaluate their method's fairness on a film dataset and a micro-loan dataset. The protected item groups were selected among film genres that are unevenly recommended to different genders for the film dataset, and among the most unfunded regions in the micro-loan dataset.

\subsubsection{Preferential Calibration}
\begin{equation}
1 - \frac{\delta_u-\delta_{\text{min}\_u}}{\delta_{\text{max}\_u}-\delta_{\text{min}\_u}}\nonumber
\end{equation}
\cite{paraschakis_matchmaking_2020} consider a matchmaking scenario where users can set their preference for being matched with people belonging to the same sensitive group as them. The proposed metric measures how well the provided recommendation respects the user's preferences, a concept known as calibrated recommendation \citep{calibration_2018}. First, the optimal recommendation composition is calculated based on the provided user preference and the sensitive group composition of the full population. $\delta_u$ is then calculated as the absolute difference between the ideal and actual composition. Finally, the normalized $\delta$ is subtracted from 1 after identifying the best and worst possible $\delta$-s for the actual user, yielding a metric for which values closer to 1 indicate that the users' preferences are better respected.

\subsubsection{Intrinsic Sensitive Attribute Match}
\begin{equation}
    \text{IntrinsicSensitiveMatch}(u,v) = (\text{UserIntrinsic}_u - \text{ItemIntrinsic}_v)^2\nonumber
\end{equation}
\cite{bobadilla_deepfair_2021} devise a notion of intrinsic sensitive properties in both items and users. They assign values to the said property for items by first considering the ratio of female users who like the items compared to the ratio of female users who dislike them, then taking the difference between that value and the equivalent value calculated for male users. These item values are then used reversely to assign values to the same intrinsic properties of the individual users. The fairness of the individual recommendations is set to be the squared difference between the intrinsic user value and the intrinsic item value. 

\subsubsection{Sensitive Neutral Recommended Items}
\begin{align}
    \text{IF}(u) &= \sum_{v \in \text{Rec}_u} \text{SensitiveEntropy}_{v}\nonumber\\
    \text{DIF}(u) & = \text{IF}(u) - \text{IF}_\text{ground\_truth}(u),\nonumber
\end{align}
Here $\text{SensitiveEntropy}_v$ is the information entropy of the sensitive attribute of the users involved in the interactions with the item $v$. The information entropy of an item is maximized when different sensitive groups historically have interacted with it at an identical rate, which is considered ideal in the fairness view of \cite{li_fairsr_2022}. The full metric is the difference between the summed entropy of the recommended items and the summed entropy of the \textit{ground truth} recommendations, which can be interpreted as how much more neutral are the recommended items compared with the \textit{correct} recommendations. 

\subsubsection{Segment Recommendation Frequency Parity}
\cite{wan_addressing_2020} considers a segmented market, where each segment covers a group of users and a group of products. Within this setting, they argue that the distribution of recommendations given across segments should match the observed data across the same segments. To that end, they construct frequency distributions that represent how the segments are represented in the observations and the recommendations and calculate a distance, i.e., unfairness, between the two using \textit{Kullback-Leibler} divergence \textit{of} the recommended frequencies \textit{from} the observed frequencies. 

\subsubsection{Sensitive Reclassification, Pre-/Post-}
Analogous to the Neutral Representation metric Sensitive Reclassification, some studies measure how well sensitive attributes can be identified using the input or output data. The pre-processing approach of \cite{slokom_towards_2021} reports the AUC achieved by auxiliary classifiers tasked with identifying sensitive attributes given user data modified by their approach. Similarly, the in-processing approach of \cite{wu_fairness-aware_2021}, and the post-processing approach of \cite{edizel_fairecsys_2020} collectively report accuracy, macro-F1 and \textit{Balanced Error Rate}(BER) achieved by analogous classifiers that are fed recommendation sets. BER is the sum of the false positive rate and the false negative rate divided by two.

\section{Datasets}\label{sec:data}
Like all machine learning models, recommender systems rely heavily on the datasets used to train them, i.e., the recommender systems are optimized to capture the correlations that can be observed in the datasets. The datasets are also pivotal in evaluating and comparing different approaches, and can highlight how well the approaches perform, scale, and generalize. While there is no lack of actors that could benefit from recommender systems and who possess vast amounts of user data to train models on, the sensitive nature of user data often limits the viability, or even legality, of sharing the data publicly. The sensitive nature of the data is further enhanced if sensitive attributes of the users are included, which is required when training many fairness-aware approaches. Furthermore, high-profile examples demonstrating that anonymization of data may not suffice in protecting the identity of the users \citep{deanon_2008}, along with an increasing focus on user protection in international discourse and legislation, have likely further deterred actors from sharing their data.

There is a conspicuous lack of relevant datasets for evaluating consumer-side fairness in recommender systems. This discrepancy is both in terms of the total number of available datasets and the relevancy of the domains they represent. The ethical ramifications of discriminatory recommender systems are better highlighted by a career recommendation setting than through a movie recommendation setting. While it is not unlikely that learned social roles partly explain current differences in the movie preferences of male and female consumers, the further propagation of such preferences is arguably less severe than consistently recommending low-income careers to female career seekers because of biases in the data. 

An overview of all datasets that contain information on sensitive attributes can be found in Table \ref{tab:datsets_group} and covers the context of the datasets, the presence of a selection of consumer-side sensitive attributes and a tally of the number of studies that have evaluated their models on the specific datasets. The MovieLens datasets\citep{movielens} dominate the field. Among the variations of the dataset, the 1-million and the 100-thousand alone contain sensitive consumer attributes in the form of gender, age and occupation. A wide adoption of the same dataset poses numerous benefits, like improved comparisons and calibrations. However, one ideally wants multiple widely adapted datasets, as different datasets usually pose different challenges, and good performance on one dataset does not necessitate good general performance. Eight studies considers various datasets based on LastFM, who all share domain but vary in size, scope and time of collection. The second most adapted \textit{singular} dataset applied for consumer-side fairness is the Google Local Review dataset\citep{google_local}, yet it is only considered by a total of four different studies, neither of which consider the MovieLens dataset. Of the remaining datasets, only a handful are used for evaluation in more than a single study, and many of these only appear more than once because they are applied in multiple studies by the same research group. For instance, three of the four studies using the Google Local Review datasets share a subset of authors. It is safe to say that the field currently lacks an established set of benchmark datasets.

Regarding the domains covered by the different datasets, most cover the recommendation of digital media, commodities or services. The few datasets that present more sensitive scenarios have not managed to attract attention in a field that is starved for relevant data, which may imply other limiting factors like restricted availability, dataset size and data quality. In particular, the aforementioned privacy concerns likely play an essential role in the lack of relevant datasets.

When factoring in dataset occurrence counts, most studies consider datasets that provide information on the consumer's gender, age and occupation. Among these, gender is the most widely adopted sensitive attribute and is typically split into two groups, male and female. The adoption of age as a sensitive attribute is also prevalent, and the attribute has been split into two or more groups based on age intervals. Occupation is rarely used, which has been attributed to difficulties related to the high number of highly skewed labels that make empiric evaluation difficult and possibly misleading. 

The datasets listed in Table \ref{tab:datsets_indv} do not explicitly provide sensitive information and have either been used to evaluate individual fairness or have supplemented such information using other means. For instance, \cite{bose_compositional_2019} compiled a dataset using the Reddit API \citep{reddit_2022}, only comprising users who have explicitly provided gender in communities they partake in that require this. \cite{paraschakis_matchmaking_2020} consider matchmaking and use demographic data on the religious adherence of different races in the US to probabilistically model a ``same religion" attribute for linking with an explicitly provided ``same religion preference" attribute. Finally, \cite{fang_fairroad_2022} derived gender labels of Yelp users \cite{yelp_2022} based on their provided names.

\begin{table}[!htb]
\tiny
\begin{center}
\caption{Table representing the different datasets applied in fair consumer-side recommender systems research with sensitive attributes.}\label{tab:datsets_group}%
\begin{tabular}{p{0.1\textwidth}p{0.1\textwidth}p{0.08\textwidth}p{0.02\textwidth}p{0.02\textwidth}p{0.02\textwidth}p{0.02\textwidth}p{0.02\textwidth}p{0.02\textwidth}p{0.02\textwidth}p{0.02\textwidth}p{0.02\textwidth}p{0.02\textwidth}p{0.02\textwidth}}
\toprule
 &  &  & \multicolumn{6}{@{}c@{}}{\textbf{Sensitive attribute}} & & \multicolumn{4}{@{}c@{}}{\textbf{Count}}\\\cmidrule{4-9}\cmidrule{11-14}
 & \rotatebox[origin=l]{90}{\textbf{Reference}} & \rotatebox[origin=l]{90}{\textbf{Setting}} & \rotatebox[origin=l]{90}{\textbf{Gender}} & \rotatebox[origin=l]{90}{\textbf{Age}} & \rotatebox[origin=l]{90}{\textbf{Marital Status}} & \rotatebox[origin=l]{90}{\textbf{Occupation}} & \rotatebox[origin=l]{90}{\textbf{Race/Nationality}} & \rotatebox[origin=l]{90}{\textbf{Region}} & \rotatebox[origin=l]{90}{\textbf{Body Size}} & \rotatebox[origin=l]{90}{\textbf{Pre-}} & \rotatebox[origin=l]{90}{\textbf{In-}} & \rotatebox[origin=l]{90}{\textbf{Post-}} & \rotatebox[origin=l]{90}{\textbf{Total}}\\
\midrule
\textbf{MovieLens} & \cite{movielens} & Films & \checkmark & \checkmark & & \checkmark & & & & 2 & 22 & 3 & 27\\
& & & & & & & & & & & & &\\
\textbf{LastFM}\footnotemark[1] & \cite{lastfm_2022} & Music & \checkmark & \checkmark & & \checkmark & & & & 1 & 3 & 4 & 8\\
& & & & & & & & & & & & &\\
\textbf{FourSquare} & \cite{foursquare_2021} & Locations & \checkmark & &  &  &  & & & 0 & 3 & 0 & 3\\
& & & & & & & & & & & & &\\
\textbf{IJCAI2015} & \cite{ijcai_2022} & Shopping &  & \checkmark &  &  & & & & 0 & 3 & 0 & 3\\
& & & & & & & & & & & & &\\
\textbf{Amazon Electronic} & \cite{wan_addressing_2020} & Electronic Articles & \checkmark &  &  &  & & & & 0 & 3 & 0 & 3\\
& & & & & & & & & & & & &\\
\textbf{Sushi} & \cite{sushi_2022} & Sushi & \checkmark & \checkmark & & & & & & 0 & 2 & 0 & 2 \\
& & & & & & & & & & & & &\\
\textbf{Speeddate} & \cite{speeddate_2006} & Dating Matches & & & & & \checkmark & & & 0 & 1 & 1 & 2\\
& & & & & & & & & & & & &\\
\textbf{Facebook}\footnotemark[3] & \cite{facebook_2015} & College Majors & \checkmark & & & & & & & 0 & 2 & 0 & 2\\
& & & & & & & & & & & & &\\
\textbf{Book-Crossing} & \cite{book_crossing_2005} & Books & \checkmark & \checkmark &  &  & & & & 0 & 2 & 0 & 2\\
& & & & & & & & & & & & &\\
\textbf{Kiva}\footnotemark[1] & \cite{kiva_2022} & Micro-lending & & & & & & \checkmark\footnotemark[2] & & 0 & 1 & 0 & 1\\
& & & & & & & & & & & & &\\
\textbf{Twitter, expert topic} & \cite{expert_topic_2016} & Experts + Topics & & & & & \checkmark\footnotemark[2] & & & 0 & 1 & 0 & 1\\
& & & & & & & & & & & & &\\
\textbf{DBLP} & \cite{dblp_2008} & Co-authors & & & & & & \checkmark\footnotemark[2] & & 0 & 1 & 0 & 1\\
& & & & & & & & & & & & &\\
\textbf{Insurance} & \cite{insurance_2022} & Insurance Products & \checkmark & & \checkmark & \checkmark & & & & 0 & 1 & 0 & 1\\
& & & & & & & & & & & & &\\
\textbf{ModCloth} & \cite{wan_addressing_2020} & Clothing &  &  &  &  & & & \checkmark & 0 & 1 & 0 & 1\\
& & & & & & & & & & & & &\\
\textbf{MSN News} & \cite{msn_news_2019} & News & \checkmark & &  &  & & & & 0 & 1 & 0 & 1\\
& & & & & & & & & & & & &\\
\textbf{Instagram} & \cite{instagram_2018} & Locations & \checkmark & \checkmark &  &  & & & & 0 & 1 & 0 & 1\\
& & & & & & & & & & & & &\\
\textbf{MathNation}\footnotemark[4] & \cite{mathnation_2022} & Learning Peers & \checkmark & &  &  & \checkmark & & & 0 & 1 & 0 & 1\\
& & & & & & & & & & & & &\\
\textbf{CIKM 2019} & \cite{cikm_2022} & E-commerce & \checkmark & \checkmark &  &  & & & & 0 & 1 & 0 & 1\\
& & & & & & & & & & & & &\\
\textbf{Taobao Ad} & \cite{taobao_2022} & Ads & \checkmark & \checkmark &  &  & & & & 0 & 1 & 0 & 1\\
\bottomrule
\end{tabular}
\end{center}
\end{table}

\begin{table}[htb]
\tiny
\begin{center}
\caption{Table representing the different datasets applied in fair consumer-side recommender systems research, without sensitive attributes.}\label{tab:datsets_indv}%
\begin{tabular}{p{0.15\textwidth}p{0.25\textwidth}p{0.15\textwidth}p{0.05\textwidth}p{0.05\textwidth}p{0.05\textwidth}p{0.05\textwidth}}
\toprule
 &  &  &  \multicolumn{4}{@{}c@{}}{\textbf{Count}}\\\cmidrule{4-7}
 & \textbf{Reference} & \textbf{Setting} & \textbf{Pre-} & \textbf{In-} & \textbf{Post-} & \textbf{Total}\\
\midrule
\textbf{Google Local Review} & \cite{google_local} & Locations & 0 & 0 & 4 & 4\\
& & & & & & \\
& & & & & & \\
\textbf{Flixster} & \cite{flixter_2010} & Films & 1 & 2 & 0 & 3\\
& & & & & & \\
\textbf{Reddit}\footnotemark[1] & \cite{reddit_2022} & Forum Boards & 0 & 1 & 1 & 2\\
& & & & & & \\
\textbf{Amazon} & \cite{amazon_2016} & E-commerce & 0 & 1 & 1 & 2\\
& & & & & & \\
\textbf{Yelp Challenge} & \cite{yelp_2022}\footnotemark[2] & Locations & 1 & 0 & 0 & 1\\
& & & & & & \\
\textbf{Freebase15k-237} & \cite{freebase_2015} & Knowledge Base Completion & 0 & 1 & 0 & 1\\
& & & & & & \\
\textbf{BeerAdvocate} & \cite{beeradvocate_2012} & Beers & 0 & 1 & 0 & 1\\
& & & & & & \\
\textbf{DPG Recruitment}\footnotemark[3] & \cite{dpg_2022} & Jobs & 0 & 1 & 0 & 1\\
& & & & & & \\
\textbf{Twitter, scientific rumour} & \cite{DeDomenico2013} & Followers & 0 & 0 & 1 & 1\\
& & & & & & \\
\textbf{Ctrip}\footnotemark[3] & \cite{ctrip_2022} & Flights & 0 & 0 & 1 & 1\\
\bottomrule
\end{tabular}
\end{center}
\end{table}

\section{Future Directions}
Given how new the field is, it is not easy to identify and recommend promising directions in terms of model architecture, etc. There is likely also not a single fairness definition that everyone will be able to agree on, so the field is undoubtedly going to continue exploring multiple parallel directions in the foreseeable future. However, regarding reproducibility aspects and other measures for improving the credibility of the research and approaches, various points could benefit from additional focus in the coming years. In particular, we perceive a need for consolidating fairness concepts, working towards standardizing the fairness metrics and improving comparisons with other approaches.
\footnotetext[1]{The service offers an API that has been used to compile different datasets.}
\footnotetext[2]{The sensitive attribute relates to the provider side, not the consumer side.}
\footnotetext[3]{The dataset is no longer distributed.}
\footnotetext[4]{The dataset is not publicly available.}

\subsection{Consolidated Fairness Concepts and Definitions}
\footnotetext[1]{The service offers an API that has been used to compile different datasets.}
\footnotetext[2]{The variation of the dataset used in the Yelp challenge and by \cite{fang_fairroad_2022}, is no longer distributed.}
\footnotetext[3]{The dataset is not publicly available.}
A recurring observation in the studies covered in this survey is the lack of a common language when it comes to fairness concepts and definitions. It often falls to the reader to interpret exactly what the authors consider fair by examining the text, the implementation choices and the evaluation. This survey highlights that there are multiple fairness definitions researched that differ significantly on a conceptual level and that are often conflicting in terms of optimization and goals. These factors complicate the examination of new research as well as comparisons of different models, and a common understanding of high-level fairness concepts could do much in to remedy such challenges. One may enhance the reader's ability to put new approaches, as well as implementation and evaluation choices, into context by immediately and accurately conveying the high-level fairness interpretation. In this case, the readers do not have to fully grasp the finer details and implications of the specific interpretation before they are able to make sense of the discussion and draw parallels with approaches they are familiar with. This may also assist researchers in identifying relevant research, and help structure further research while leaving room for more specific formal definitions within the high-level interpretations. The Fairness Interpretations taxonomy proposed in Section \ref{sec:fair_cat} is one suggestion for such high-level conceptual categories.

\subsection{Consensus on Fairness Metrics}
Section \ref{sec:metrics} demonstrates a great number of applied fairness metrics and a high degree of overlap in what they essentially seek to measure. While this is natural for a budding field, and enhanced by the presence of multiple distinct and conflicting fairness definitions, it is currently a contributing factor in making the comparisons challenging. Guided by rigorous analysis of the properties of different metrics, the field as a whole could benefit from reducing the number of metrics applied by identifying the best among metrics that have higher degrees of overlap.

\subsection{Comparison}
Despite a growing number of studies covering similar fairness concepts, there is still a low degree of comparative analysis of different approaches. While it is interesting to see how fairness-aware contributions affect the fairness over the base recommender approaches, it is also essential to compare with relevant fairness-aware approaches, if present. This aspect seems to have improved recently, but there is still room for further improvement. 

One contributing factor to the lack of comparative analysis is likely visibility. The research field is still relatively new, and the nomenclature has yet to consolidate, making it challenging to identify similar research. There is also an issue of visibility across different types of approaches, in particular recommender systems, IR Ranking and link-prediction. Both IR Ranking and link-prediction approaches may be considered recommender systems, depending on the setting or the applied dataset. However, since they use different terms than those used in the recommender system research and intermingling between fields can be uncommon, such approaches may not be known by researchers proposing similar recommender systems. Visibility has also been limited so far by the lack of updated surveys that chart out the field's current state. However, recent contributions like the comparative analysis in \cite{stvg_reprod_2022} and future surveys will hopefully improve this aspect.

\subsection{Datasets}
As noted in Section \ref{sec:data}, there are currently not that many relevant datasets for evaluating consumer-side recommender systems fairness. A wider selection of benchmarking datasets could improve evaluation and comparisons and add credibility to the research. New datasets should ideally vary in size and sources to offer different challenges related to aspects like scalability and adaptability, focusing on filling in the gaps not covered by the datasets applied today. In particular, many current datasets are getting old, and their application may fail to reflect performance in a shifting environment. Finally, to better highlight the need and application of fair recommender systems, it would be useful to have datasets for which the ethical implications of a discriminatory recommender system are more severe.

\subsection{Online Evaluation}
None of the considered studies performs an online evaluation of neither recommendation utility nor fairness. While offline evaluation has some practical benefits, it is usually restricted to only being able to reward recommendation of items/actions we know the user likes, not serendipitous recommendations of items the user will like but was not aware of when the dataset was created. Online A/B testing, on the other hand, can reward such recommendations and may bring along other benefits of testing the model in the environment it will be used, granted that they are designed and executed well. Further, online evaluation allows more subjective feedback, e.g., asking the users if they suspect that the recommender system discriminates against them or presents them with biased recommendations influenced by their inferred or stated sensitive attributes.

While researchers like \cite{impression_fairness_2019} looks into the public pre-conceived perception and attitude towards formal fairness definitions, the impression of those using a fairness-aware recommender system may differ. Multiple approaches covered in this survey strive to make their models or recommendations independent of sensitive attributes. It would be interesting to see how different users perceive such a system in different recommendation settings.



\section*{Declarations}
\subsection*{Funding}
This publication has been partly funded by the SFI NorwAI, (Centre for Research-based Innovation, 309834). The authors gratefully acknowledge the financial support from the Research Council of Norway and the partners of the SFI NorwAI.

\bibliographystyle{sn-basic}
\bibliography{main.bib}

\end{document}